\documentclass[
 twocolumn,superscriptaddress,aps,prl,amsmath,nofootinbib,amssymb,floatfix]{revtex4-2}
\usepackage{graphicx}
\usepackage{dcolumn}
\usepackage{bm}
\usepackage{dsfont}
\usepackage[breaklinks=true,colorlinks,citecolor=blue,linkcolor=blue,urlcolor=blue]{hyperref}

\usepackage{physics}
\usepackage{mleftright}
\usepackage{bbm}

\newcommand{\be}{\begin{equation}}
\newcommand{\ee}{\end{equation}}
\newcommand{\bea}{\begin{eqnarray}}
\newcommand{\eea}{\end{eqnarray}}

\renewcommand{\eqref}[1]{\mbox{Eq.~(\ref{#1})}}

\newcommand{\figpanel}[2]{Fig.~\hyperref[#1]{\ref*{#1}(#2)}}

\usepackage{changes}
\definechangesauthor[name={Ale}, color=red]{Ale}
\definechangesauthor[name={David}, color=orange]{David}

\usepackage{tikz,xcolor}
\definecolor{lime}{HTML}{A6CE39}
\DeclareRobustCommand{\orcidicon}{%
	\begin{tikzpicture}
	\draw[lime, fill=lime] (0,0) 
	circle [radius=0.16] 
	node[white] {{\fontfamily{qag}\selectfont \tiny ID}};
	\draw[white, fill=white] (-0.0625,0.095) 
	circle [radius=0.007];
	\end{tikzpicture}
	\hspace{-2mm}
}
\foreach \x in {A, ..., Z}{%
	\expandafter\xdef\csname orcid\x\endcsname{\noexpand\href{https://orcid.org/\csname orcidauthor\x\endcsname}{\noexpand\orcidicon}}
}

\begin{document}
\title{Quantum field heat engine powered by phonon-photon interactions
}

\date{\today}

\author{Alessandro Ferreri\orcidA{}}
\affiliation{Institute for Quantum Computing Analytics (PGI-12), Forschungszentrum J\"ulich, 52425 J\"ulich, Germany}
\affiliation{Theoretical Quantum Physics Laboratory, RIKEN, Wako-shi, Saitama 351-0198, Japan}
\author{Vincenzo Macrì\orcidE{}}
\affiliation{Theoretical Quantum Physics Laboratory, RIKEN, Wako-shi, Saitama 351-0198, Japan}
\affiliation{Dipartimento di Ingegneria, Universit\`{a} degli Studi di Palermo, Viale delle Scienze, 90128 Palermo, Italy}
\author{Frank K. Wilhelm\orcidC{}}
\affiliation{Institute for Quantum Computing Analytics (PGI-12), Forschungszentrum J\"ulich, 52425 J\"ulich, Germany}
\author{Franco Nori\orcidD{}}
\affiliation{Theoretical Quantum Physics Laboratory, RIKEN, Wako-shi, Saitama 351-0198, Japan}
\affiliation{Center for Quantum Computing, RIKEN, Wako-shi, Saitama 351-0198, Japan}
\affiliation{Physics Department, The University of Michigan, Ann Arbor, Michigan 48109-1040, USA}
\author{David Edward Bruschi\orcidB{}}
\affiliation{Institute for Quantum Computing Analytics (PGI-12), Forschungszentrum J\"ulich, 52425 J\"ulich, Germany}

\begin{abstract}
We present a quantum heat engine based on a cavity with two oscillating mirrors that confine a quantum field as the working substance. The engine performs an Otto cycle during which the walls and a field mode interact via a nonlinear Hamiltonian.
Resonances between the frequencies of the cavity mode and the walls allow to transfer heat from the hot and the cold bath by exploiting the conversion between phononic and photonic excitations. We study the time evolution of the system and show that net work can be extracted after a full cycle. We evaluate the efficiency of the process. 
\end{abstract}


\maketitle

{\emph{Introduction.}}---
Quantum thermodynamics studies physical processes at the quantum scale through the lens of thermodynamics  \cite{gemmer2009quantum,Vinjanampathy2016, kurizki_kofman_2022}.
The overall aim of this field of research is to extend concepts initially developed in the classical theory, such as heat, work and thermodynamic efficiency,  into the quantum domain where purely quantum features can be exploited \cite{skrzypczyk_work_2014, pekola_towards_2015}. These include quantum correlations \cite{perarnau-llobet_extractable_2015}, quantum coherence \cite{scully_extracting_2003} and vacuum fluctuations \cite{Ezzahri2014,Pendry2016}. 
One task is to propose and characterize novel thermodynamic cycles by taking advantage of the nonclassical nature of the working substance to extract work for different tasks \cite{quan_quantum_2007, quan_quantum_2009}. The interest in studying thermodynamical cycles at the quantum level goes beyond the mere possibility of reaching higher degrees of efficiency but aims towards miniaturization of future thermal machines based on quantum systems.

The quantum Otto cycle is a thermodynamic cycle that enjoys relative ease of theoretical implementation in a quantum framework  compared to other cycles, especially if implemented in cavity-optomechanics \cite{zhang_quantum_2014, zhang_theory_2014, mari_quantum_2015,  brunelli_out--equilibrium_2015, kolar_extracting_2016,PhysRevLett.124.210601}. It consists of a combination of two thermodynamic ``strokes": 
i) the \textit{isochoric transformation}, performed by maintaining constant the spacing of the energy levels of the system during thermalization with the bath; ii) the \textit{adiabatic} transformation, where the thermally isolated system evolves with the number of excitations kept constant.
This simple cycle has been implemented and studied in the context of finite-time Otto cycles \cite{PhysRevE.103.022136, PhysRevE.100.042126, PhysRevResearch.2.033083}, Otto-engine power generation \cite{PhysRevE.100.062140, PhysRevE.102.062123, campisi_power_2016}, heat engines with interacting systems \cite{PhysRevResearch.5.013088}, and quantum heat engines based on phononic fields in Bose-Einstein condensates \cite{PRXQuantum.2.030310}.

In this work we study the quantum Otto cycle in the context of quantum optomechanics \cite{DiStefano2019}. The system consists of a cavity-optomechanical setup where two movable mirrors  confine a quantum field. The mirror and field modes strongly interact via phonon-photon vacuum fluctuations, and the mirrors are also coupled individually to a thermal bath. We employ our system to show that the proposed quantum field heat engine can generate power in finite time after each cycle, and we estimate the efficiency of such process.
We note that the platforms considered here have already allowed for the experimental observation of thermal-phonon hopping, i.e, the exchange of thermal energy between individual phonon modes \cite{fong_phonon_2019}.

In contrast to the standard quantum Otto cycle, the cavity mode here plays the role of working substance and does not interact with the  baths directly. Instead, the interaction is mediated by the quantized position degree of freedom of the respective wall. More precisely, each wall is connected to a thermal bath: the first wall (called W1) interacts with a \textit{cold} bath at temperature $T_{\textrm{c}}$, while the second wall (called W2) interacts with a \textit{hot} bath at temperature $T_{\textrm{h}}$. The transfer of thermal excitations from the bath to the cavity mode occurs whenever the frequency of the cavity is resonant with the frequency of the corresponding wall. The field modes are driven by an external drive, which controls the length of the cavity.  

We stress that the photon-phonon interaction occurs beyond linearization, thereby retaining the dynamical-Casimir-like three-body terms in the Hamiltonian \cite{law_interaction_1995,butera_field_2013,macri_nonperturbative_2018,PhysRevA.106.033502, PhysRevA.100.062501, dodonov_fifty_2020}, with the ambition of characterizing the thermodynamic performance of the system in its full nonlinear regime. This includes also the multimode character of the interaction \cite{bruschi_time_2019}.
Moreover, in contrast to standard studies of four-stroke thermal machines that consider the single strokes separately, we employed the master equation formalism in order to run two consecutive cycles as a function of time. The need for the master equation approach throughout the whole dynamics is explained by the fact that we effectively change the length of the cavity in a time-dependent way through the external drive, thereby controlling the heat transfer between the cavity and the baths through the walls. 

\begin{figure}[ht!]
	\centering
	\includegraphics[width=0.9\linewidth]{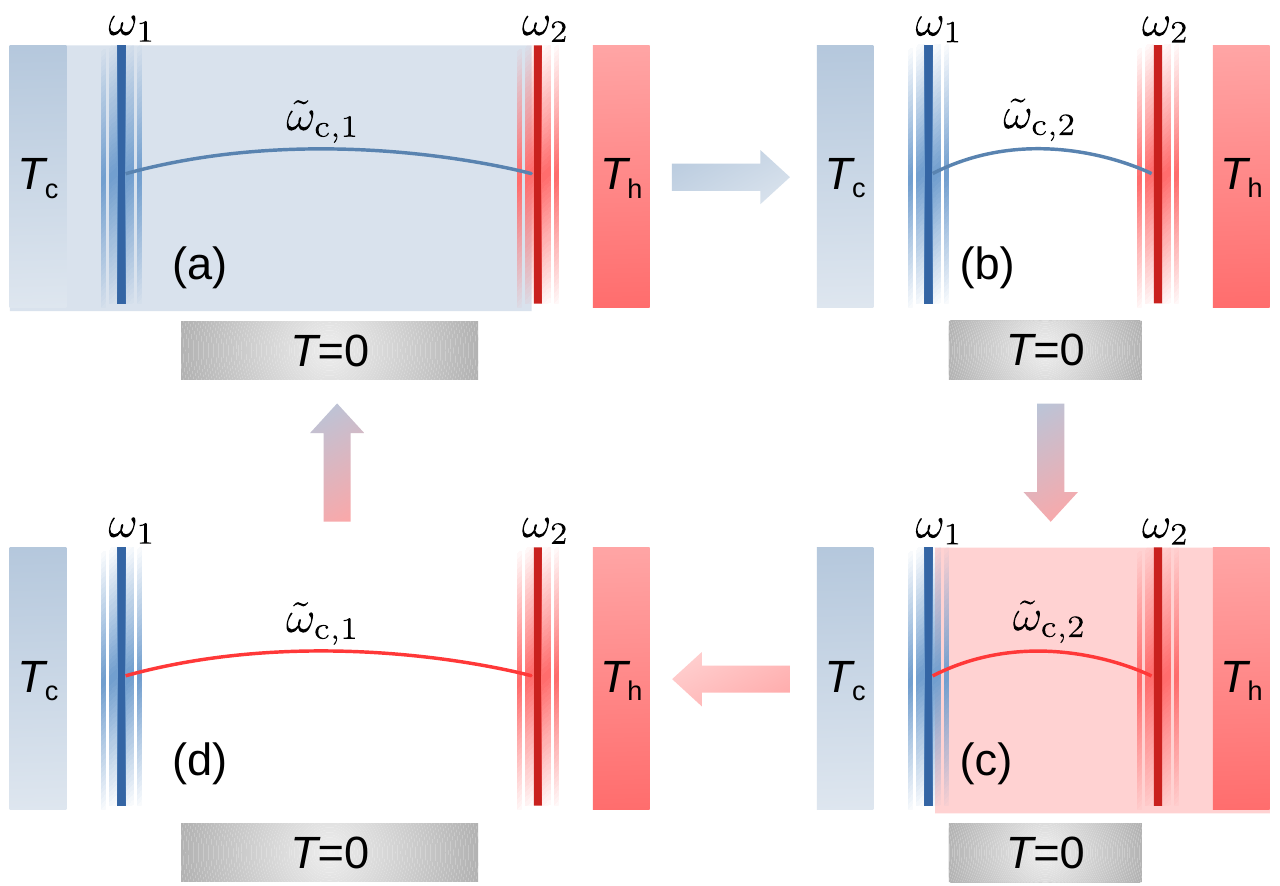}
	\caption{Pictorial representation of the cavity performing the quantum Otto cycle: two movable walls, coupled to two baths at different temperatures, confine a cavity mode, which is weakly coupled to a bath at $T\simeq 0$. The four panels describe the four strokes of the cycle: (a) cold isochoric, (b) adiabatic compression, (c) hot isochoric, (d) adiabatic expansion.}
	\label{cycle}
\end{figure}

{\emph{The quantum model.}}--- The system is composed of two movable walls that confine an uncharged massless scalar quantum field and interact with a local bath. Our choice of field is a good approximation for a single-polarization version of a confined electromagnetic spin-1 field \cite{Friis:Lee:2013}. We further simplify the setup by considering a one dimensional cavity, which allows us to obtain the system Hamiltonian $\hat H_s$ following the standard procedure of solving the classical field equations and then quantizing \cite{PhysRevA.106.033502}. The system is schematically represented in Fig.\ref{cycle}.

The Hamiltonian can be split as usual as $\hat H_\textrm{s}=\hat H_0+\hat H_\textrm{I}$, 
where $\hat H_0=\omega_1\hat b_1^\dag\hat b_1+\omega_2\hat b_2^\dag\hat b_2+\omega_{\textrm{c}}\hat a^\dag\hat a$ is the bare Hamiltonian and $H_I$ is the interaction Hamiltonian. The latter has been previously obtained \cite{DiStefano2019}, and reads
\begin{align}
\hat H_\textrm{I}=&\frac{g_1}{2}(\hat a+\hat a^\dag)^2(\hat b_1+\hat b_1^\dag)+\frac{g_2}{2}(\hat a+\hat a^\dag)^2(\hat b_2+\hat b_2^\dag)\label{hi}.
\end{align}
Here, $\hat{a},\hat{a}^\dag$ are the photonic operators for the cavity-field mode, while $\hat{b}_j,\hat{b}^\dag_j$ ($j=1,2$) are the phononic operators for the walls-field mode. The operators satisfy the canonical commutator relations $[\hat{a},\hat{a}^\dag]=1$ and 
$[\hat{b}_j,\hat{b}^\dag_{j'}]=\delta_{jj'}$. 
Furthermore, $\omega_j$ are the frequencies of the two  movable walls ($\omega_1<\omega_2$ for convenience), whereas $\omega_{\textrm{c}}$ is the frequency of the cavity field mode. In addition, the coupling constants $g_j$ quantify the strength of the coupling between the field mode and the $j$th-wall (see Appendix). Finally, eigenvalues and states will be labelled by $l,m,n\in\mathbb{N}$, which stand for $l$ excitations of the field mode, $m$ excitations of W1 and excitations $n$ of W2. Throughout this work we assume that $\hbar=c=k_B=1$.

The interaction Hamiltonian \eqref{hi} contains three types of terms: (i) The radiation pressure $\hat a^\dag\hat a(\hat b_j+\hat b_j^\dag)$, paramount in standard optomechanics \cite{aspelmeyer2012quantum, Aspelmeyer2014}, which shifts the cavity frequency $\omega_{\textrm{c}}$ in case of coherent motion of the wall. (ii) The excitation transfer terms $\hat a^2\hat b_j^\dag+(\hat a^\dag)^2\hat b_j$, which convert single-phonon excitations into photon pairs (and vice versa). In other words, they convert mechanical and electromagnetic energy into each other. In order for the photons to appear and contribute during the dynamics, resonance conditions $k\omega_j=2\omega_{\textrm{c}}$ with $k\in \mathbb{N}$ involving high-frequency movable walls must be fulfilled \cite{macri_nonperturbative_2018}. Such resonances can be achieved in current optomechanical setups using real movable mirrors \cite{Connell2010quantum}. However, they play the most significant role in experimental platforms based on superconducting circuits \cite{Johansson2010,Wilson2011,johansson_optomechanical-like_2014,Kim2015}.
(iii) The counter-rotating terms $\hat a^2\hat b_j+(\hat a^\dag)^2\hat b_j^\dag$, which generally contribute to modifying the energy-density of the cavity field (because of the quantum wall fluctuations \cite{butera_field_2013}) and are also responsible for the non-conservation of the particle number. Counter-rotating terms are involved in virtual processes that would allow to observe higher-order coherent processes in cavity-optomechanics \cite{Wang2023,Russo2023}. 
 
\begin{figure}[ht!]
	\centering
	\includegraphics[width=0.9\linewidth]{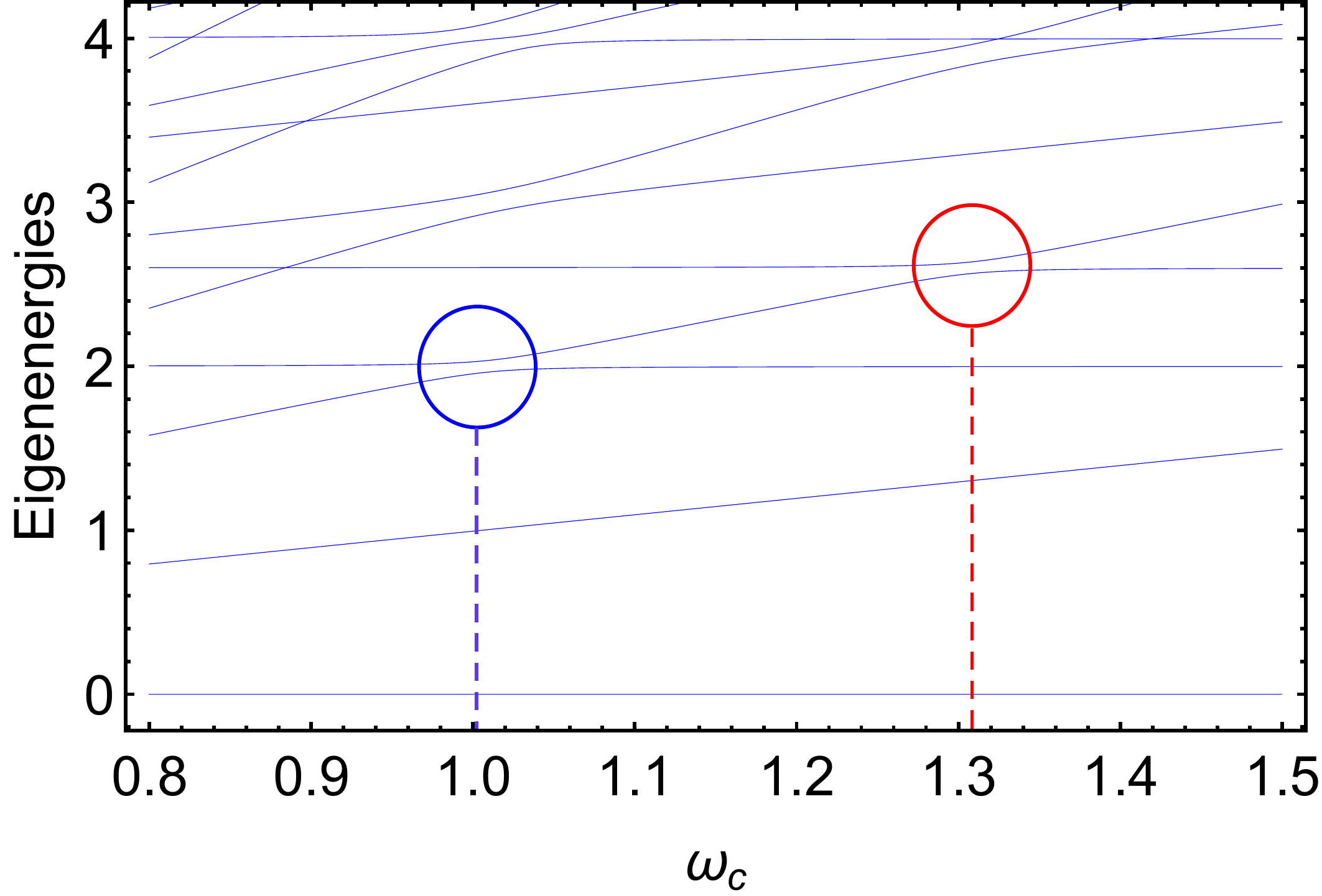}
	\caption{Lowest-energy levels of the system Hamiltonian as a function of $\omega_{\textrm{c}}$. Circles highlight the avoided levels in which the dressed resonances (blue) $\omega_1=2\omega_{\textrm{c},1}$ and (red) $\omega_2=2\omega_{\textrm{c},2}$ occur. For each resonance condition, the effective cavity frequency $\tilde\omega_{\textrm{c}}$ was estimated. Here: $\tilde\omega_{\textrm{c},1}=1.01$, $\tilde\omega_{\textrm{c},2}=1.31$, $\omega_1=2$, $\omega_2=2.6$, $g_1=g_2=0.05$. Frequencies are normalized with respect to the bare frequency of the cavity mode $\omega_{\textrm{c},1}$.}
	\label{al}
\end{figure}

The interaction between the bosonic modes not only inevitably alters the structure of the energy levels with respect to the bare ones, but it also lifts the degeneracy in the presence of resonances. 
To this end we diagonalize the Hamiltonian $H_\textrm{s}$ numerically and plot the energy levels in Fig.~\ref{al} for different values of the frequency $\omega_{\textrm{c}}$. The figure clearly shows the presence of avoided level crossings due to the energy split in proximity of the frequency values $\omega_{\textrm{c},1}=\omega_1/2$ and $\omega_{\textrm{c},2}=\omega_2/2$, i.e., where the resonances are expected to occur as discussed above. 
The dashed vertical lines in Fig.~\ref{al} highlight the shift in the bare frequencies.
The effective cavity-frequencies $\tilde\omega_{\textrm{c},i}$ ($i=1,2$) involved in the coherent resonant processes can be estimated numerically by calculating the minimal splitting of the avoided level crossings.  

We use open quantum system dynamics to compute all quantities of interest \cite{breuer2002theory}. This requires us to solve the master equation for the density operator $\hat{\rho}$ representing the state. Since we are considering a strongly interacting system, we employ the tools developed in the literature \cite{beaudoin_dissipation_2011,settineri2018,PhysRevA.106.033502}. In particular, we employ the Lindblad equation $\dot {\hat{\rho}}=-i[\hat H,\hat{\rho}]+{\hat{\mathcal{L}}_{\rm D}}\hat{\rho}$, where ${\hat{\mathcal{L}}_{\rm D}}$ indicates the Lindblad superoperator expressed in the dressed base (see Appendix).
Here we assume that the three subsystems are coupled to three different baths: W1 is coupled to a cold bath with damping rate $\gamma_1$ and temperature $T_{\textrm{c}}$; W2 is coupled to a hot bath with damping rate $\gamma_2$ and temperature $T_{\textrm{h}}$; the cavity mode interacts with its own bath with damping rate $\kappa\simeq 0$ temperature $T\simeq 0$. This is a reasonable assumption for realistic implementations.

{\emph{Quantum Otto cycle.}}---The main idea of this work is to \textit{perform the quantum Otto cycle by using the two walls as bosonic channels to perform the heat transfer between the hot and cold baths through the cavity mode by exploiting the mechanical-electromagnetic energy conversion}. We present here the framework employed to achieve our goal. 

Any heat engine can be characterized by evaluating the output power $\mathcal{P}$ and the efficiency $\eta$ of a single thermodynamic cycle.
Thus, we consider the total Hamiltonian $\hat H_{\textrm{tot}}(t)=\hat H_\textrm{s}+\hat H_{\textrm{dr}}(t)$ that includes the time-dependent drive term $\hat H_{\textrm{dr}}(t)$.
This contribution is expressed in the dressed picture and periodically drives the cavity frequency from $\tilde\omega_{\textrm{c},1}$ to $\tilde\omega_{\textrm{c},2}$ and vice versa, simulating the physical process of compression and expansion of the cavity. This can be understood from the fact that the frequency of a trapped field mode decreases or increases by respectively enhancing or reducing the length of the cavity. Thus, changes in frequency simulate changes in length. Concretely, we have 
$\hat H_{\textrm{dr}}(t)=f(t) \Delta\omega\hat A^\dag\hat A$, where $\Delta\omega=\tilde\omega_{\textrm{c},2}-\tilde\omega_{\textrm{c},1}$, $\hat A,\hat A^\dag$ are the cavity dressed operators obtained by diagonalizing the Hamiltonian, and $f(t)$ is a periodic smooth step function (see Appendix). 
Such step functions are commonly employed in circuit quantum electrodynamics (QED) \cite{sillanpaa2007coherent,hofheinz2008generation,hofheinz2009synthesizing,Wang2011,mariantoni2011photon,kockum2017frequency}. Furthermore, time dependent drives have also been considered in studies of nonlinear optomechanics \cite{PhysRevA.104.013501, PhysRevA.101.033834, Qvarfort_2020}.

In order to investigate the thermodynamic features of the system, we define the change of internal energy $\Delta \mathcal{U}(t)=\textrm{Tr}[\hat H_{\textrm{tot}}(t)\rho(t)]-\textrm{Tr}[\hat H_{\textrm{tot}}(0)\rho(0)]$, the change in heat $\Delta\mathcal{Q}(t)=\int_0^t d\tau \,\textrm{Tr}\bigl[\hat H_{\textrm{tot}}(\tau)\dot{\rho}(\tau)\bigr]$, and the change in work $\Delta\mathcal{W}(t)=\int_0^t d\tau \,\textrm{Tr}\left[\dot{\hat H}_{\textrm{tot}}(\tau)\rho(\tau)\right]$. These three quantities satisfy the first law of thermodynamics in its quantum formulation:
\begin{align}
\Delta \mathcal{U}(t)= \Delta\mathcal{Q}(t)+\Delta\mathcal{W}(t).
\label{fl}
\end{align}
We then provide a formal definition for the output power $\mathcal{P}$ and the efficiency $\eta$ of a thermodynamical process as
\begin{align}\label{power:efficiency}
    \mathcal{P}:= \frac{d}{dt}\Delta\mathcal{W}(t), \quad \eta:=-\frac{\mathcal{W}^{\textrm{out}}}{\mathcal{Q}^{\textrm{in}}},
\end{align}
where $\mathcal{W}^{\textrm{out}}$ and $\mathcal{Q}^{\textrm{in}}$ are, respectively, the work provided and the heat absorbed by the system.
These are the main expressions evaluated in our work.


We now present our quantum Otto cycle, which is composed of a preliminary phase and four steps: (0) \textit{Initialization}: the whole system is prepared in its vacuum state $\rho(0)=\lvert 0\rangle\langle 0\rvert$ and we assume that the cavity is initially coupled to W1 by means of the resonance condition $\omega_{\textrm{c},1}=\omega_1/2$ (see Fig.~\ref{cycle}); (a) \textit{ Cold isochoric}: Thermal cold phonons from W1 are converted into photons, until the subsystem W1+cavity is thermalized; (b) \textit{Adiabatic compression}: The external drive shifts the field frequency from $\tilde\omega_{\textrm{c},1}$ to $\tilde\omega_{\textrm{c},2}$ ; (c) \textit{ Hot isochoric}: The newly activated resonance $\omega_{\textrm{c}2}=\omega_2/2$ facilitates the excitation transfer between the cavity mode and W2, during which hot phonons are converted into photons; (d)  \textit{ Adiabatic expansion}: The drive changes the cavity mode frequency back to the resonance regime $\omega_{\textrm{c},1}=\omega_1/2$. After this step, the system is ready to restart from the \textit{cold isochoric} stroke of step (a). 

We now offer a few remarks on the properties of our cycle and the differences with the standard quantum Otto cycle already known in the literature \cite{ quan_quantum_2007}. Recall that a \textit{quantum isochoric transformation} occurs whenever the population of the system changes while thermalizing with a bath and maintaining the energy level spacing constant, while a \textit{quantum adiabatic transformation} occurs whenever both the energy level spacing and the temperature are tuned such that the population does not vary throughout the process \cite{quan_quantum_2005, quan_quantum_2007}.
The requirements for a transformation to be called ``quantum adiabatic" dramatically differ with respect to the classical case. Indeed, while a classical adiabatic transformation simply forces the system to be thermally isolated throughout the process, namely not coupled to any bath, the condition of quantum adiabaticity additionally requires the stability of the population \cite{quan_quantum_2007}.
In this sense, the latter is conceptually more limiting and challenging to realize in an experiment.
Furthermore, any quantum adiabatic transformation has to be performed very slowly, namely without generating any output power  \cite{PhysRevE.100.062140, PhysRevE.102.062123, Albash_2012}.

Importantly, since the frequency change occurs much faster than the effective coupling,
the system does not have time to modify its eigenstates, which implies that we do not need to diagonalize the Hamiltonian at all times. Therefore, we can maintain the same eigenbasis, hence the same dressed operators, throughout the whole dynamics.
Moreover, the interaction between the cavity and each wall occurs not only at the minimum point of the level avoidance but also weakly in its proximity. It follows that this rapid jump between $\tilde\omega_{\textrm{c},1}$ and $\tilde\omega_{\textrm{c},2}$ is necessary to ensure the classical adiabaticity of the process \cite{quan_quantum_2007}, namely simulating the deactivation of the interaction between the cavity and the bath mediated by the wall.
In this quantum description, the two walls are thermalized via different baths. In particular, the interaction between a specific bath and the cavity occurs once the resonance condition with the relative wall is imposed, and it is halted once the system is driven off-resonance. We have W2 at frequency $\omega_2$ that thermalizes via a hot bath at $T_{\textrm{h}}$, while W1 has a lower frequency $\omega_1<\omega_2$ and it is thermalized via a cold bath at $T_{\textrm{c}}<T_{\textrm{h}}$. The cavity interacts with its own bath at $T\simeq 0$.

\begin{figure}[ht!]
	\centering
	\includegraphics[width=0.9\linewidth]{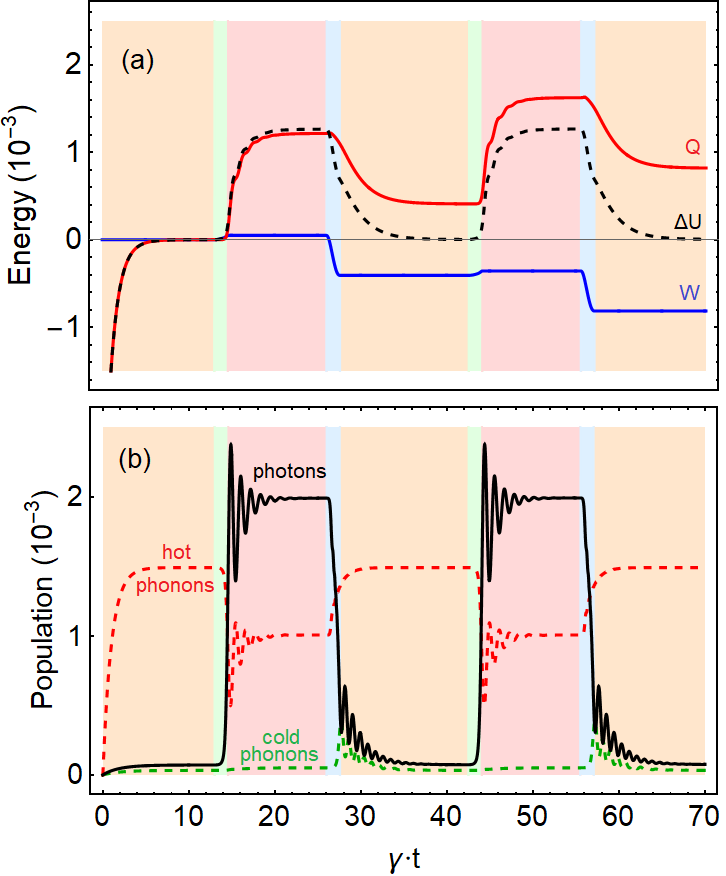}
	\caption{Time evolution of the quantities of interest. (a): variation of internal energy $\Delta \mathcal{U}$ (dashed, black), the work $\mathcal{W}$ (blue), and the heat $\mathcal{Q}$ (red). (b): population of the cavity photon (black), the cold wall 1 phonons (dashed green) and  the hot wall 2 phonons (dashed red).
 The background colors indicate the four stokes shown in Fig.~\ref{cycle}: cold isochoric (orange), adiabatic compression (green), hot isochoric (red), adiabatic expansion (blue). Here: $\omega_1=2$, $\omega_2=2.6$, $\tilde\omega_{\textrm{c},1}=1.01$, $\tilde\omega_{\textrm{c},2}=1.31$, $g_1=g_2=0.05$, $T_{\textrm{c}}=0.15$, $T_{\textrm{h}}=0.40$, $\gamma_1=\gamma_2=\gamma=0.01$, $\kappa=10^{-6}$, $T_{0}=10^{-7}$. Frequencies and temperatures are normalized with respect to $\omega_{\textrm{c},1}$.}
	\label{ener}
\end{figure}

{\emph{Numerical analysis.}}---
We have solved numerically the master equation and therefore calculated the time evolution of the average internal energy $\Delta\mathcal{U}(t)$ directly, while we have computed the average work $\Delta\mathcal{W}(t)$ by integrating the expression of the power $\mathcal{P}(t)\equiv\textrm{Tr}[\rho(t)\dot H_{\textrm{tot}}(t)]=\textrm{Tr}[\rho(t)\dot H_{\textrm{dr}}(t)]=\dot f(t) \Delta\omega\textrm{Tr}[\rho(t)\hat A^\dag\hat A]$ defined in \eqref{power:efficiency}. The average heat change $\Delta\mathcal{Q}(t)$ can be easily derived employing the first law of thermodinamics in \eqref{fl}. We have used the parameters indicated in Fig.~\ref{ener}, and our choice of frequencies has followed two important criteria: we need to clearly distinguish the two avoided levels in order to perform the jump, but at the same time we must avoid any degeneracy with other possible resonances in order to prevent unwanted heat flows that could reduce the efficiency. Our results can be found in Fig.~\ref{ener}.

We now discuss our findings. Once the dynamics start, a transient phase occurs in which both walls absorb thermal excitations from their own baths. However, while the interaction between W1 and the cavity converts part of the thermal phonons into photons, W2, which is momentarily not interacting with the cavity, thermalizes. In Fig.~\ref{ener}a, $\Delta \mathcal{U}=0$ corresponds to the energy of the system at the end of this transient phase. Recall that thermalization of the cavity while interacting with W1 is defined as stroke (a).
During stroke (b), i.e., during compression, the cavity absorbs work from the drive and increases $\tilde\omega_{\textrm{c},1}$ to $\tilde\omega_{\textrm{c},2}$, eventually starting to absorb thermal excitations from the hot wall and converting them into photons [which defines stroke (c)]. This causes a drastic enhancement of the photon population at the expense of part of the phonon population, as seen in Fig.~\ref{ener}b. At the same time, the population of the cold wall, now off-resonance, increases and W1 thermalizes completely. Once the internal energy becomes constant and the populations stop fluctuating, we perform the rapid expansion of the cavity as described in stroke (d), which causes the system to release an amount of energy which is higher than the one initially absorbed, a key feature of a properly functioning heat engine. This net gain becomes evident by looking at $\Delta\mathcal{W}$ in Fig.~\ref{ener}a. After this last stroke, the cavity thermalizes with W1 again and the system reaches its initial configuration: a new cycle can now take place. We conclude that, after each cycle, we can extract a net amount of work using our system.

At this point, we wish to provide a measure of quality for the proposed cycle.
This can be done by means of the efficiency defined in \eqref{power:efficiency}, which we compute using the parameters employed in the figures. We find an efficiency of $\eta$=0.376, which can be compared to the Carnot limit $\eta_C=1-T_{\textrm{c}}/T_{\textrm{h}}$=0.625 as a benchmark for the performance of the machine. Further benchmarking could be carried out by evaluating the efficiency at maximum power \cite{Albash_2012}. Such analysis would require a detailed numerical optimization of this efficiency as a function of the frequencies of the two walls, and it could include the activation of further resonances during the adiabatic transformations.
This is beyond the scope of the present work, and it is left for future investigations.

We now provide a few remarks on our proposal. As mentioned before we note that, once the resonance with the wall $j$ is implemented, the interaction between wall $j$ and the cavity mode is given by terms $(\hat a^\dag)^2\hat b_j+\hat a^2\hat b_j^\dag$ in $\hat H_\textrm{I}$, which mediate the phonon-photon conversions. From a physical point of view this means that the degree of freedom that thermalizes during any isochoric stroke is \textit{not} the cavity mode alone, but is instead the \textit{combination} of the cavity mode \textit{and} the activated wall. In other words, during the isochoric transformation the change in photon population is not a direct consequence of the thermalization of the cavity mode with the relative bath, instead it is a consequence of the \textit{partial conversion of thermal phonons into photon pairs}. This process continues until the system wall-cavity reaches the steady state, as can be observed in Fig.~\ref{ener}b. Thus, the number of photons at the end of any isochoric process will be less than $N=(e^{\omega_{\textrm{c}}/T_0}-1)^{-1}$ as predicted by Bose-Einstein statistics and expected in the Otto cycle with a single-mode as working substance \cite{abah_single-ion_2012, rosnagel_nanoscale_2014}. In particular, the photon population in Fig.\ref{ener}b can be compared with the average number of photons for a single cavity mode with frequency $\tilde\omega_{\textrm{c},1}=1.01$ and temperature $T_{\textrm{c}}=0.15$, which is $N=1.2\cdot 10^{-3}$, and the population of a single-cavity mode with frequency $\tilde\omega_{\textrm{c},2}=1.31$ and temperature $T_{\textrm{h}}=0.4$, which is $N=3.96\cdot 10^{-2}$. It can be seen that the photon population in our system is about 94\% less than what expected by a direct thermalization with the bath  at the end of the cold isochoric, and about 95\% less at the end of the hot isochoric. This suggests that we can run our machine with fewer resources. 

Furthermore, we notice that, 
during the isochoric strokes,
quantum excitations are extracted from the thermal fluctuations of the walls and converted into photons \textit{without requiring the walls to perform any classical motion}. This is in contrast to the semiclassical description of our setup, wherein walls are treated as classical degrees of freedom \cite{dodonov_fifty_2020}. 
The phonon-photon conversion occurs between two quantum channels by simply activating the resonance $\omega_i=2\omega_{\textrm{c},i}$ ($i=1,2$).
The possibility of extracting thermal excitations from the walls in case of no coherent motion is a consequence of the fact that the two mirrors operate in the quantum regime  \cite{PhysRevA.100.022501, PhysRevA.106.033502}.

Finally, we stress that, in contrast to other works studying optomechanical systems in a quantum thermodynamic framework \cite{zhang_quantum_2014, zhang_theory_2014}, we employ a nonlinear Hamiltonian that includes the nonlinear interaction between the movable walls and the electromagnetic mode. 
In particular, the entire dynamics of our system is based on the conversion of phonons into photons and vice versa. To our knowledge, this effect has never been considered in a quantum thermodynamic context.
This highlights the importance of moving beyond the linearized regime.

{\emph{Conclusions.}}---In this work we proposed a quantum heat engine based on a cavity system composed of a scalar field trapped by two  fluctuating walls that performs an Otto cycle. In our setup, we exploit the phonon-photon conversion mechanism to let the working substance, namely the cavity mode, exchange heat with the thermal baths during the cycle. We demonstrated that it is possible to extract net work using carefully modulated resonances, and we have evaluated the overall efficiency of the cycle. We believe that this work opens the way to the systematic study of quantum field thermodynamic engines, to be used for fundamental science, as well as the development of novel quantum technologies.

{\emph{Acknowledgments.}}---A.F. thanks the research center RIKEN for the hospitality. A.F. acknowledges the ``JSPS Summer Program 2022'' and the ``FY2022 JSPS Postdoctoral Fellowship for Research in Japan (Short-term)'', sponsored by the Japanese Society for the Promotion of Science (JSPS).
F.K.W., A.F.,
and D.E.B. acknowledge support from the joint project No. 13N15685 ``German Quantum Computer based on Superconducting Qubits (GeQCoS)'' sponsored by the German Federal Ministry of Education and Research (BMBF) under the framework “Quantum technologies–from basic research to the market”. 
F.N. is supported in part by: Nippon Telegraph and Telephone Corporation (NTT) Research, the Japan Science and Technology Agency (JST) [via the Quantum Leap Flagship Program (Q-LEAP), and the Moonshot R\&D Grant Number JPMJMS2061], the Asian Office of Aerospace Research and Development (AOARD) (via Grant No. FA2386-20-1-4069), and the Foundational Questions Institute Fund (FQXi) via Grant No. FQXi-IAF19-06.
\bibliography{alessandria}

\begin{thebibliography}{62}%
\makeatletter
\providecommand \@ifxundefined [1]{%
 \@ifx{#1\undefined}
}%
\providecommand \@ifnum [1]{%
 \ifnum #1\expandafter \@firstoftwo
 \else \expandafter \@secondoftwo
 \fi
}%
\providecommand \@ifx [1]{%
 \ifx #1\expandafter \@firstoftwo
 \else \expandafter \@secondoftwo
 \fi
}%
\providecommand \natexlab [1]{#1}%
\providecommand \enquote  [1]{``#1''}%
\providecommand \bibnamefont  [1]{#1}%
\providecommand \bibfnamefont [1]{#1}%
\providecommand \citenamefont [1]{#1}%
\providecommand \href@noop [0]{\@secondoftwo}%
\providecommand \href [0]{\begingroup \@sanitize@url \@href}%
\providecommand \@href[1]{\@@startlink{#1}\@@href}%
\providecommand \@@href[1]{\endgroup#1\@@endlink}%
\providecommand \@sanitize@url [0]{\catcode `\\12\catcode `\$12\catcode
  `\&12\catcode `\#12\catcode `\^12\catcode `\_12\catcode `\%12\relax}%
\providecommand \@@startlink[1]{}%
\providecommand \@@endlink[0]{}%
\providecommand \url  [0]{\begingroup\@sanitize@url \@url }%
\providecommand \@url [1]{\endgroup\@href {#1}{\urlprefix }}%
\providecommand \urlprefix  [0]{URL }%
\providecommand \Eprint [0]{\href }%
\providecommand \doibase [0]{https://doi.org/}%
\providecommand \selectlanguage [0]{\@gobble}%
\providecommand \bibinfo  [0]{\@secondoftwo}%
\providecommand \bibfield  [0]{\@secondoftwo}%
\providecommand \translation [1]{[#1]}%
\providecommand \BibitemOpen [0]{}%
\providecommand \bibitemStop [0]{}%
\providecommand \bibitemNoStop [0]{.\EOS\space}%
\providecommand \EOS [0]{\spacefactor3000\relax}%
\providecommand \BibitemShut  [1]{\csname bibitem#1\endcsname}%
\let\auto@bib@innerbib\@empty
\bibitem [{\citenamefont {Gemmer}\ \emph {et~al.}(2009)\citenamefont {Gemmer},
  \citenamefont {Michel},\ and\ \citenamefont {Mahler}}]{gemmer2009quantum}%
  \BibitemOpen
  \bibfield  {author} {\bibinfo {author} {\bibfnamefont {J.}~\bibnamefont
  {Gemmer}}, \bibinfo {author} {\bibfnamefont {M.}~\bibnamefont {Michel}},\
  and\ \bibinfo {author} {\bibfnamefont {G.}~\bibnamefont {Mahler}},\
  }\href@noop {} {\emph {\bibinfo {title} {Quantum thermodynamics: Emergence of
  thermodynamic behavior within composite quantum systems}}},\ Vol.\ \bibinfo
  {volume} {784}\ (\bibinfo  {publisher} {Springer},\ \bibinfo {year}
  {2009})\BibitemShut {NoStop}%
\bibitem [{\citenamefont {Vinjanampathy}\ and\ \citenamefont
  {Anders}(2016)}]{Vinjanampathy2016}%
  \BibitemOpen
  \bibfield  {author} {\bibinfo {author} {\bibfnamefont {S.}~\bibnamefont
  {Vinjanampathy}}\ and\ \bibinfo {author} {\bibfnamefont {J.}~\bibnamefont
  {Anders}},\ }\bibfield  {title} {\bibinfo {title} {{Quantum
  thermodynamics}},\ }\href {https://doi.org/10.1080/00107514.2016.1201896}
  {\bibfield  {journal} {\bibinfo  {journal} {Contemp. Phys.}\ }\textbf
  {\bibinfo {volume} {57}},\ \bibinfo {pages} {545} (\bibinfo {year}
  {2016})}\BibitemShut {NoStop}%
\bibitem [{\citenamefont {Kurizki}\ and\ \citenamefont
  {Kofman}(2022)}]{kurizki_kofman_2022}%
  \BibitemOpen
  \bibfield  {author} {\bibinfo {author} {\bibfnamefont {G.}~\bibnamefont
  {Kurizki}}\ and\ \bibinfo {author} {\bibfnamefont {A.~G.}\ \bibnamefont
  {Kofman}},\ }\href@noop {} {\emph {\bibinfo {title} {Thermodynamics and
  Control of Open Quantum Systems}}}\ (\bibinfo  {publisher} {Cambridge
  University Press},\ \bibinfo {year} {2022})\BibitemShut {NoStop}%
\bibitem [{\citenamefont {Skrzypczyk}\ \emph {et~al.}(2014)\citenamefont
  {Skrzypczyk}, \citenamefont {Short},\ and\ \citenamefont
  {Popescu}}]{skrzypczyk_work_2014}%
  \BibitemOpen
  \bibfield  {author} {\bibinfo {author} {\bibfnamefont {P.}~\bibnamefont
  {Skrzypczyk}}, \bibinfo {author} {\bibfnamefont {A.~J.}\ \bibnamefont
  {Short}},\ and\ \bibinfo {author} {\bibfnamefont {S.}~\bibnamefont
  {Popescu}},\ }\bibfield  {title} {\bibinfo {title} {Work extraction and
  thermodynamics for individual quantum systems},\ }\href
  {https://doi.org/10.1038/ncomms5185} {\bibfield  {journal} {\bibinfo
  {journal} {Nature Communications}\ }\textbf {\bibinfo {volume} {5}},\
  \bibinfo {pages} {4185} (\bibinfo {year} {2014})}\BibitemShut {NoStop}%
\bibitem [{\citenamefont {Pekola}(2015)}]{pekola_towards_2015}%
  \BibitemOpen
  \bibfield  {author} {\bibinfo {author} {\bibfnamefont {J.~P.}\ \bibnamefont
  {Pekola}},\ }\bibfield  {title} {\bibinfo {title} {Towards quantum
  thermodynamics in electronic circuits},\ }\href
  {https://doi.org/10.1038/nphys3169} {\bibfield  {journal} {\bibinfo
  {journal} {Nature Physics}\ }\textbf {\bibinfo {volume} {11}},\ \bibinfo
  {pages} {118} (\bibinfo {year} {2015})}\BibitemShut {NoStop}%
\bibitem [{\citenamefont {Perarnau-Llobet}\ \emph {et~al.}(2015)\citenamefont
  {Perarnau-Llobet}, \citenamefont {Hovhannisyan}, \citenamefont {Huber},
  \citenamefont {Skrzypczyk}, \citenamefont {Brunner},\ and\ \citenamefont
  {Acín}}]{perarnau-llobet_extractable_2015}%
  \BibitemOpen
  \bibfield  {author} {\bibinfo {author} {\bibfnamefont {M.}~\bibnamefont
  {Perarnau-Llobet}}, \bibinfo {author} {\bibfnamefont {K.~V.}\ \bibnamefont
  {Hovhannisyan}}, \bibinfo {author} {\bibfnamefont {M.}~\bibnamefont {Huber}},
  \bibinfo {author} {\bibfnamefont {P.}~\bibnamefont {Skrzypczyk}}, \bibinfo
  {author} {\bibfnamefont {N.}~\bibnamefont {Brunner}},\ and\ \bibinfo {author}
  {\bibfnamefont {A.}~\bibnamefont {Acín}},\ }\bibfield  {title} {\bibinfo
  {title} {Extractable {Work} from {Correlations}},\ }\href
  {https://doi.org/10.1103/PhysRevX.5.041011} {\bibfield  {journal} {\bibinfo
  {journal} {Physical Review X}\ }\textbf {\bibinfo {volume} {5}},\ \bibinfo
  {pages} {041011} (\bibinfo {year} {2015})}\BibitemShut {NoStop}%
\bibitem [{\citenamefont {Scully}\ \emph {et~al.}(2003)\citenamefont {Scully},
  \citenamefont {Zubairy}, \citenamefont {Agarwal},\ and\ \citenamefont
  {Walther}}]{scully_extracting_2003}%
  \BibitemOpen
  \bibfield  {author} {\bibinfo {author} {\bibfnamefont {M.~O.}\ \bibnamefont
  {Scully}}, \bibinfo {author} {\bibfnamefont {M.~S.}\ \bibnamefont {Zubairy}},
  \bibinfo {author} {\bibfnamefont {G.~S.}\ \bibnamefont {Agarwal}},\ and\
  \bibinfo {author} {\bibfnamefont {H.}~\bibnamefont {Walther}},\ }\bibfield
  {title} {\bibinfo {title} {Extracting {Work} from a {Single} {Heat} {Bath}
  via {Vanishing} {Quantum} {Coherence}},\ }\href
  {https://doi.org/10.1126/science.1078955} {\bibfield  {journal} {\bibinfo
  {journal} {Science}\ }\textbf {\bibinfo {volume} {299}},\ \bibinfo {pages}
  {862} (\bibinfo {year} {2003})}\BibitemShut {NoStop}%
\bibitem [{\citenamefont {Ezzahri}\ and\ \citenamefont
  {Joulain}(2014)}]{Ezzahri2014}%
  \BibitemOpen
  \bibfield  {author} {\bibinfo {author} {\bibfnamefont {Y.}~\bibnamefont
  {Ezzahri}}\ and\ \bibinfo {author} {\bibfnamefont {K.}~\bibnamefont
  {Joulain}},\ }\bibfield  {title} {\bibinfo {title} {Vacuum-induced phonon
  transfer between two solid dielectric materials: Illustrating the case of
  casimir force coupling},\ }\href {https://doi.org/10.1103/PhysRevB.90.115433}
  {\bibfield  {journal} {\bibinfo  {journal} {Phys. Rev. B}\ }\textbf {\bibinfo
  {volume} {90}},\ \bibinfo {pages} {115433} (\bibinfo {year}
  {2014})}\BibitemShut {NoStop}%
\bibitem [{\citenamefont {Pendry}\ \emph {et~al.}(2016)\citenamefont {Pendry},
  \citenamefont {Sasihithlu},\ and\ \citenamefont {Craster}}]{Pendry2016}%
  \BibitemOpen
  \bibfield  {author} {\bibinfo {author} {\bibfnamefont {J.~B.}\ \bibnamefont
  {Pendry}}, \bibinfo {author} {\bibfnamefont {K.}~\bibnamefont {Sasihithlu}},\
  and\ \bibinfo {author} {\bibfnamefont {R.~V.}\ \bibnamefont {Craster}},\
  }\bibfield  {title} {\bibinfo {title} {Phonon-assisted heat transfer between
  vacuum-separated surfaces},\ }\href
  {https://doi.org/10.1103/PhysRevB.94.075414} {\bibfield  {journal} {\bibinfo
  {journal} {Phys. Rev. B}\ }\textbf {\bibinfo {volume} {94}},\ \bibinfo
  {pages} {075414} (\bibinfo {year} {2016})}\BibitemShut {NoStop}%
\bibitem [{\citenamefont {Quan}\ \emph {et~al.}(2007)\citenamefont {Quan},
  \citenamefont {Liu}, \citenamefont {Sun},\ and\ \citenamefont
  {Nori}}]{quan_quantum_2007}%
  \BibitemOpen
  \bibfield  {author} {\bibinfo {author} {\bibfnamefont {H.~T.}\ \bibnamefont
  {Quan}}, \bibinfo {author} {\bibfnamefont {Y.-x.}\ \bibnamefont {Liu}},
  \bibinfo {author} {\bibfnamefont {C.~P.}\ \bibnamefont {Sun}},\ and\ \bibinfo
  {author} {\bibfnamefont {F.}~\bibnamefont {Nori}},\ }\bibfield  {title}
  {\bibinfo {title} {Quantum thermodynamic cycles and quantum heat engines},\
  }\href {https://doi.org/10.1103/PhysRevE.76.031105} {\bibfield  {journal}
  {\bibinfo  {journal} {Physical Review E}\ }\textbf {\bibinfo {volume} {76}},\
  \bibinfo {pages} {031105} (\bibinfo {year} {2007})}\BibitemShut {NoStop}%
\bibitem [{\citenamefont {Quan}(2009)}]{quan_quantum_2009}%
  \BibitemOpen
  \bibfield  {author} {\bibinfo {author} {\bibfnamefont {H.~T.}\ \bibnamefont
  {Quan}},\ }\bibfield  {title} {\bibinfo {title} {Quantum thermodynamic cycles
  and quantum heat engines. {II}.},\ }\href
  {https://doi.org/10.1103/PhysRevE.79.041129} {\bibfield  {journal} {\bibinfo
  {journal} {Physical Review E}\ }\textbf {\bibinfo {volume} {79}},\ \bibinfo
  {pages} {041129} (\bibinfo {year} {2009})}\BibitemShut {NoStop}%
\bibitem [{\citenamefont {Zhang}\ \emph
  {et~al.}(2014{\natexlab{a}})\citenamefont {Zhang}, \citenamefont {Bariani},\
  and\ \citenamefont {Meystre}}]{zhang_quantum_2014}%
  \BibitemOpen
  \bibfield  {author} {\bibinfo {author} {\bibfnamefont {K.}~\bibnamefont
  {Zhang}}, \bibinfo {author} {\bibfnamefont {F.}~\bibnamefont {Bariani}},\
  and\ \bibinfo {author} {\bibfnamefont {P.}~\bibnamefont {Meystre}},\
  }\bibfield  {title} {\bibinfo {title} {Quantum {Optomechanical} {Heat}
  {Engine}},\ }\href {https://doi.org/10.1103/PhysRevLett.112.150602}
  {\bibfield  {journal} {\bibinfo  {journal} {Physical Review Letters}\
  }\textbf {\bibinfo {volume} {112}},\ \bibinfo {pages} {150602} (\bibinfo
  {year} {2014}{\natexlab{a}})}\BibitemShut {NoStop}%
\bibitem [{\citenamefont {Zhang}\ \emph
  {et~al.}(2014{\natexlab{b}})\citenamefont {Zhang}, \citenamefont {Bariani},\
  and\ \citenamefont {Meystre}}]{zhang_theory_2014}%
  \BibitemOpen
  \bibfield  {author} {\bibinfo {author} {\bibfnamefont {K.}~\bibnamefont
  {Zhang}}, \bibinfo {author} {\bibfnamefont {F.}~\bibnamefont {Bariani}},\
  and\ \bibinfo {author} {\bibfnamefont {P.}~\bibnamefont {Meystre}},\
  }\bibfield  {title} {\bibinfo {title} {Theory of an optomechanical quantum
  heat engine},\ }\href {https://doi.org/10.1103/PhysRevA.90.023819} {\bibfield
   {journal} {\bibinfo  {journal} {Physical Review A}\ }\textbf {\bibinfo
  {volume} {90}},\ \bibinfo {pages} {023819} (\bibinfo {year}
  {2014}{\natexlab{b}})}\BibitemShut {NoStop}%
\bibitem [{\citenamefont {Mari}\ \emph {et~al.}(2015)\citenamefont {Mari},
  \citenamefont {Farace},\ and\ \citenamefont
  {Giovannetti}}]{mari_quantum_2015}%
  \BibitemOpen
  \bibfield  {author} {\bibinfo {author} {\bibfnamefont {A.}~\bibnamefont
  {Mari}}, \bibinfo {author} {\bibfnamefont {A.}~\bibnamefont {Farace}},\ and\
  \bibinfo {author} {\bibfnamefont {V.}~\bibnamefont {Giovannetti}},\
  }\bibfield  {title} {\bibinfo {title} {Quantum optomechanical piston engines
  powered by heat},\ }\href {https://doi.org/10.1088/0953-4075/48/17/175501}
  {\bibfield  {journal} {\bibinfo  {journal} {Journal of Physics B: Atomic,
  Molecular and Optical Physics}\ }\textbf {\bibinfo {volume} {48}},\ \bibinfo
  {pages} {175501} (\bibinfo {year} {2015})}\BibitemShut {NoStop}%
\bibitem [{\citenamefont {Brunelli}\ \emph {et~al.}(2015)\citenamefont
  {Brunelli}, \citenamefont {Xuereb}, \citenamefont {Ferraro}, \citenamefont
  {Chiara}, \citenamefont {Kiesel},\ and\ \citenamefont
  {Paternostro}}]{brunelli_out--equilibrium_2015}%
  \BibitemOpen
  \bibfield  {author} {\bibinfo {author} {\bibfnamefont {M.}~\bibnamefont
  {Brunelli}}, \bibinfo {author} {\bibfnamefont {A.}~\bibnamefont {Xuereb}},
  \bibinfo {author} {\bibfnamefont {A.}~\bibnamefont {Ferraro}}, \bibinfo
  {author} {\bibfnamefont {G.~D.}\ \bibnamefont {Chiara}}, \bibinfo {author}
  {\bibfnamefont {N.}~\bibnamefont {Kiesel}},\ and\ \bibinfo {author}
  {\bibfnamefont {M.}~\bibnamefont {Paternostro}},\ }\bibfield  {title}
  {\bibinfo {title} {Out-of-equilibrium thermodynamics of quantum
  optomechanical systems},\ }\href
  {https://doi.org/10.1088/1367-2630/17/3/035016} {\bibfield  {journal}
  {\bibinfo  {journal} {New Journal of Physics}\ }\textbf {\bibinfo {volume}
  {17}},\ \bibinfo {pages} {035016} (\bibinfo {year} {2015})}\BibitemShut
  {NoStop}%
\bibitem [{\citenamefont {Kolář}\ \emph {et~al.}(2016)\citenamefont
  {Kolář}, \citenamefont {Ryabov},\ and\ \citenamefont
  {Filip}}]{kolar_extracting_2016}%
  \BibitemOpen
  \bibfield  {author} {\bibinfo {author} {\bibfnamefont {M.}~\bibnamefont
  {Kolář}}, \bibinfo {author} {\bibfnamefont {A.}~\bibnamefont {Ryabov}},\
  and\ \bibinfo {author} {\bibfnamefont {R.}~\bibnamefont {Filip}},\ }\bibfield
   {title} {\bibinfo {title} {Extracting work from quantum states of
  radiation},\ }\href {https://doi.org/10.1103/PhysRevA.93.063822} {\bibfield
  {journal} {\bibinfo  {journal} {Physical Review A}\ }\textbf {\bibinfo
  {volume} {93}},\ \bibinfo {pages} {063822} (\bibinfo {year}
  {2016})}\BibitemShut {NoStop}%
\bibitem [{\citenamefont {Holmes}\ \emph {et~al.}(2020)\citenamefont {Holmes},
  \citenamefont {Anders},\ and\ \citenamefont
  {Mintert}}]{PhysRevLett.124.210601}%
  \BibitemOpen
  \bibfield  {author} {\bibinfo {author} {\bibfnamefont {Z.}~\bibnamefont
  {Holmes}}, \bibinfo {author} {\bibfnamefont {J.}~\bibnamefont {Anders}},\
  and\ \bibinfo {author} {\bibfnamefont {F.}~\bibnamefont {Mintert}},\
  }\bibfield  {title} {\bibinfo {title} {Enhanced energy transfer to an
  optomechanical piston from indistinguishable photons},\ }\href
  {https://doi.org/10.1103/PhysRevLett.124.210601} {\bibfield  {journal}
  {\bibinfo  {journal} {Phys. Rev. Lett.}\ }\textbf {\bibinfo {volume} {124}},\
  \bibinfo {pages} {210601} (\bibinfo {year} {2020})}\BibitemShut {NoStop}%
\bibitem [{\citenamefont {Lee}\ \emph {et~al.}(2021)\citenamefont {Lee},
  \citenamefont {Ha},\ and\ \citenamefont {Jeong}}]{PhysRevE.103.022136}%
  \BibitemOpen
  \bibfield  {author} {\bibinfo {author} {\bibfnamefont {S.}~\bibnamefont
  {Lee}}, \bibinfo {author} {\bibfnamefont {M.}~\bibnamefont {Ha}},\ and\
  \bibinfo {author} {\bibfnamefont {H.}~\bibnamefont {Jeong}},\ }\bibfield
  {title} {\bibinfo {title} {Quantumness and thermodynamic uncertainty relation
  of the finite-time otto cycle},\ }\href
  {https://doi.org/10.1103/PhysRevE.103.022136} {\bibfield  {journal} {\bibinfo
   {journal} {Phys. Rev. E}\ }\textbf {\bibinfo {volume} {103}},\ \bibinfo
  {pages} {022136} (\bibinfo {year} {2021})}\BibitemShut {NoStop}%
\bibitem [{\citenamefont {Kloc}\ \emph {et~al.}(2019)\citenamefont {Kloc},
  \citenamefont {Cejnar},\ and\ \citenamefont
  {Schaller}}]{PhysRevE.100.042126}%
  \BibitemOpen
  \bibfield  {author} {\bibinfo {author} {\bibfnamefont {M.}~\bibnamefont
  {Kloc}}, \bibinfo {author} {\bibfnamefont {P.}~\bibnamefont {Cejnar}},\ and\
  \bibinfo {author} {\bibfnamefont {G.}~\bibnamefont {Schaller}},\ }\bibfield
  {title} {\bibinfo {title} {Collective performance of a finite-time quantum
  otto cycle},\ }\href {https://doi.org/10.1103/PhysRevE.100.042126} {\bibfield
   {journal} {\bibinfo  {journal} {Phys. Rev. E}\ }\textbf {\bibinfo {volume}
  {100}},\ \bibinfo {pages} {042126} (\bibinfo {year} {2019})}\BibitemShut
  {NoStop}%
\bibitem [{\citenamefont {Das}\ and\ \citenamefont
  {Mukherjee}(2020)}]{PhysRevResearch.2.033083}%
  \BibitemOpen
  \bibfield  {author} {\bibinfo {author} {\bibfnamefont {A.}~\bibnamefont
  {Das}}\ and\ \bibinfo {author} {\bibfnamefont {V.}~\bibnamefont
  {Mukherjee}},\ }\bibfield  {title} {\bibinfo {title} {Quantum-enhanced
  finite-time otto cycle},\ }\href
  {https://doi.org/10.1103/PhysRevResearch.2.033083} {\bibfield  {journal}
  {\bibinfo  {journal} {Phys. Rev. Res.}\ }\textbf {\bibinfo {volume} {2}},\
  \bibinfo {pages} {033083} (\bibinfo {year} {2020})}\BibitemShut {NoStop}%
\bibitem [{\citenamefont {Chen}\ \emph {et~al.}(2019)\citenamefont {Chen},
  \citenamefont {Sun},\ and\ \citenamefont {Dong}}]{PhysRevE.100.062140}%
  \BibitemOpen
  \bibfield  {author} {\bibinfo {author} {\bibfnamefont {J.-F.}\ \bibnamefont
  {Chen}}, \bibinfo {author} {\bibfnamefont {C.-P.}\ \bibnamefont {Sun}},\ and\
  \bibinfo {author} {\bibfnamefont {H.}~\bibnamefont {Dong}},\ }\bibfield
  {title} {\bibinfo {title} {Achieve higher efficiency at maximum power with
  finite-time quantum otto cycle},\ }\href
  {https://doi.org/10.1103/PhysRevE.100.062140} {\bibfield  {journal} {\bibinfo
   {journal} {Phys. Rev. E}\ }\textbf {\bibinfo {volume} {100}},\ \bibinfo
  {pages} {062140} (\bibinfo {year} {2019})}\BibitemShut {NoStop}%
\bibitem [{\citenamefont {Singh}\ and\ \citenamefont {M\"ustecapl\ifmmode
  \imath \else \i \fi{}o\ifmmode~\breve{g}\else
  \u{g}\fi{}lu}(2020)}]{PhysRevE.102.062123}%
  \BibitemOpen
  \bibfield  {author} {\bibinfo {author} {\bibfnamefont {V.}~\bibnamefont
  {Singh}}\ and\ \bibinfo {author} {\bibfnamefont {O.~E.}\ \bibnamefont
  {M\"ustecapl\ifmmode \imath \else \i \fi{}o\ifmmode~\breve{g}\else
  \u{g}\fi{}lu}},\ }\bibfield  {title} {\bibinfo {title} {Performance bounds of
  nonadiabatic quantum harmonic otto engine and refrigerator under a squeezed
  thermal reservoir},\ }\href {https://doi.org/10.1103/PhysRevE.102.062123}
  {\bibfield  {journal} {\bibinfo  {journal} {Phys. Rev. E}\ }\textbf {\bibinfo
  {volume} {102}},\ \bibinfo {pages} {062123} (\bibinfo {year}
  {2020})}\BibitemShut {NoStop}%
\bibitem [{\citenamefont {Campisi}\ and\ \citenamefont
  {Fazio}(2016)}]{campisi_power_2016}%
  \BibitemOpen
  \bibfield  {author} {\bibinfo {author} {\bibfnamefont {M.}~\bibnamefont
  {Campisi}}\ and\ \bibinfo {author} {\bibfnamefont {R.}~\bibnamefont
  {Fazio}},\ }\bibfield  {title} {\bibinfo {title} {The power of a critical
  heat engine},\ }\href {https://doi.org/10.1038/ncomms11895} {\bibfield
  {journal} {\bibinfo  {journal} {Nature Communications}\ }\textbf {\bibinfo
  {volume} {7}},\ \bibinfo {pages} {11895} (\bibinfo {year}
  {2016})}\BibitemShut {NoStop}%
\bibitem [{\citenamefont {Boubakour}\ \emph {et~al.}(2023)\citenamefont
  {Boubakour}, \citenamefont {Fogarty},\ and\ \citenamefont
  {Busch}}]{PhysRevResearch.5.013088}%
  \BibitemOpen
  \bibfield  {author} {\bibinfo {author} {\bibfnamefont {M.}~\bibnamefont
  {Boubakour}}, \bibinfo {author} {\bibfnamefont {T.}~\bibnamefont {Fogarty}},\
  and\ \bibinfo {author} {\bibfnamefont {T.}~\bibnamefont {Busch}},\ }\bibfield
   {title} {\bibinfo {title} {Interaction-enhanced quantum heat engine},\
  }\href {https://doi.org/10.1103/PhysRevResearch.5.013088} {\bibfield
  {journal} {\bibinfo  {journal} {Phys. Rev. Res.}\ }\textbf {\bibinfo {volume}
  {5}},\ \bibinfo {pages} {013088} (\bibinfo {year} {2023})}\BibitemShut
  {NoStop}%
\bibitem [{\citenamefont {Gluza}\ \emph {et~al.}(2021)\citenamefont {Gluza},
  \citenamefont {Sabino}, \citenamefont {Ng}, \citenamefont {Vitagliano},
  \citenamefont {Pezzutto}, \citenamefont {Omar}, \citenamefont {Mazets},
  \citenamefont {Huber}, \citenamefont {Schmiedmayer},\ and\ \citenamefont
  {Eisert}}]{PRXQuantum.2.030310}%
  \BibitemOpen
  \bibfield  {author} {\bibinfo {author} {\bibfnamefont {M.}~\bibnamefont
  {Gluza}}, \bibinfo {author} {\bibfnamefont {J.~a.}\ \bibnamefont {Sabino}},
  \bibinfo {author} {\bibfnamefont {N.~H.}\ \bibnamefont {Ng}}, \bibinfo
  {author} {\bibfnamefont {G.}~\bibnamefont {Vitagliano}}, \bibinfo {author}
  {\bibfnamefont {M.}~\bibnamefont {Pezzutto}}, \bibinfo {author}
  {\bibfnamefont {Y.}~\bibnamefont {Omar}}, \bibinfo {author} {\bibfnamefont
  {I.}~\bibnamefont {Mazets}}, \bibinfo {author} {\bibfnamefont
  {M.}~\bibnamefont {Huber}}, \bibinfo {author} {\bibfnamefont
  {J.}~\bibnamefont {Schmiedmayer}},\ and\ \bibinfo {author} {\bibfnamefont
  {J.}~\bibnamefont {Eisert}},\ }\bibfield  {title} {\bibinfo {title} {Quantum
  field thermal machines},\ }\href
  {https://doi.org/10.1103/PRXQuantum.2.030310} {\bibfield  {journal} {\bibinfo
   {journal} {PRX Quantum}\ }\textbf {\bibinfo {volume} {2}},\ \bibinfo {pages}
  {030310} (\bibinfo {year} {2021})}\BibitemShut {NoStop}%
\bibitem [{\citenamefont {Di~Stefano}\ \emph {et~al.}(2019)\citenamefont
  {Di~Stefano}, \citenamefont {Settineri}, \citenamefont {Macr\`{\i}},
  \citenamefont {Ridolfo}, \citenamefont {Stassi}, \citenamefont {Kockum},
  \citenamefont {Savasta},\ and\ \citenamefont {Nori}}]{DiStefano2019}%
  \BibitemOpen
  \bibfield  {author} {\bibinfo {author} {\bibfnamefont {O.}~\bibnamefont
  {Di~Stefano}}, \bibinfo {author} {\bibfnamefont {A.}~\bibnamefont
  {Settineri}}, \bibinfo {author} {\bibfnamefont {V.}~\bibnamefont
  {Macr\`{\i}}}, \bibinfo {author} {\bibfnamefont {A.}~\bibnamefont {Ridolfo}},
  \bibinfo {author} {\bibfnamefont {R.}~\bibnamefont {Stassi}}, \bibinfo
  {author} {\bibfnamefont {A.~F.}\ \bibnamefont {Kockum}}, \bibinfo {author}
  {\bibfnamefont {S.}~\bibnamefont {Savasta}},\ and\ \bibinfo {author}
  {\bibfnamefont {F.}~\bibnamefont {Nori}},\ }\bibfield  {title} {\bibinfo
  {title} {Interaction of mechanical oscillators mediated by the exchange of
  virtual photon pairs},\ }\href
  {https://doi.org/10.1103/PhysRevLett.122.030402} {\bibfield  {journal}
  {\bibinfo  {journal} {Phys. Rev. Lett.}\ }\textbf {\bibinfo {volume} {122}},\
  \bibinfo {pages} {030402} (\bibinfo {year} {2019})}\BibitemShut {NoStop}%
\bibitem [{\citenamefont {Fong}\ \emph {et~al.}(2019)\citenamefont {Fong},
  \citenamefont {Li}, \citenamefont {Zhao}, \citenamefont {Yang}, \citenamefont
  {Wang},\ and\ \citenamefont {Zhang}}]{fong_phonon_2019}%
  \BibitemOpen
  \bibfield  {author} {\bibinfo {author} {\bibfnamefont {K.~Y.}\ \bibnamefont
  {Fong}}, \bibinfo {author} {\bibfnamefont {H.-K.}\ \bibnamefont {Li}},
  \bibinfo {author} {\bibfnamefont {R.}~\bibnamefont {Zhao}}, \bibinfo {author}
  {\bibfnamefont {S.}~\bibnamefont {Yang}}, \bibinfo {author} {\bibfnamefont
  {Y.}~\bibnamefont {Wang}},\ and\ \bibinfo {author} {\bibfnamefont
  {X.}~\bibnamefont {Zhang}},\ }\bibfield  {title} {\bibinfo {title} {Phonon
  heat transfer across a vacuum through quantum fluctuations},\ }\href
  {https://doi.org/10.1038/s41586-019-1800-4} {\bibfield  {journal} {\bibinfo
  {journal} {Nature}\ }\textbf {\bibinfo {volume} {576}},\ \bibinfo {pages}
  {243} (\bibinfo {year} {2019})}\BibitemShut {NoStop}%
\bibitem [{\citenamefont {Law}(1995)}]{law_interaction_1995}%
  \BibitemOpen
  \bibfield  {author} {\bibinfo {author} {\bibfnamefont {C.~K.}\ \bibnamefont
  {Law}},\ }\bibfield  {title} {\bibinfo {title} {Interaction between a moving
  mirror and radiation pressure: {A} {Hamiltonian} formulation},\ }\href
  {https://doi.org/10.1103/PhysRevA.51.2537} {\bibfield  {journal} {\bibinfo
  {journal} {Physical Review A}\ }\textbf {\bibinfo {volume} {51}},\ \bibinfo
  {pages} {2537} (\bibinfo {year} {1995})}\BibitemShut {NoStop}%
\bibitem [{\citenamefont {Butera}\ and\ \citenamefont
  {Passante}(2013)}]{butera_field_2013}%
  \BibitemOpen
  \bibfield  {author} {\bibinfo {author} {\bibfnamefont {S.}~\bibnamefont
  {Butera}}\ and\ \bibinfo {author} {\bibfnamefont {R.}~\bibnamefont
  {Passante}},\ }\bibfield  {title} {\bibinfo {title} {Field {Fluctuations} in
  a {One}-{Dimensional} {Cavity} with a {Mobile} {Wall}},\ }\href
  {https://doi.org/10.1103/PhysRevLett.111.060403} {\bibfield  {journal}
  {\bibinfo  {journal} {Physical Review Letters}\ }\textbf {\bibinfo {volume}
  {111}},\ \bibinfo {pages} {060403} (\bibinfo {year} {2013})}\BibitemShut
  {NoStop}%
\bibitem [{\citenamefont {Macrì}\ \emph {et~al.}(2018)\citenamefont {Macrì},
  \citenamefont {Ridolfo}, \citenamefont {Di~Stefano}, \citenamefont {Kockum},
  \citenamefont {Nori},\ and\ \citenamefont
  {Savasta}}]{macri_nonperturbative_2018}%
  \BibitemOpen
  \bibfield  {author} {\bibinfo {author} {\bibfnamefont {V.}~\bibnamefont
  {Macrì}}, \bibinfo {author} {\bibfnamefont {A.}~\bibnamefont {Ridolfo}},
  \bibinfo {author} {\bibfnamefont {O.}~\bibnamefont {Di~Stefano}}, \bibinfo
  {author} {\bibfnamefont {A.~F.}\ \bibnamefont {Kockum}}, \bibinfo {author}
  {\bibfnamefont {F.}~\bibnamefont {Nori}},\ and\ \bibinfo {author}
  {\bibfnamefont {S.}~\bibnamefont {Savasta}},\ }\bibfield  {title} {\bibinfo
  {title} {Nonperturbative {Dynamical} {Casimir} {Effect} in {Optomechanical}
  {Systems}: {Vacuum} {Casimir}-{Rabi} {Splittings}},\ }\href
  {https://doi.org/10.1103/PhysRevX.8.011031} {\bibfield  {journal} {\bibinfo
  {journal} {Physical Review X}\ }\textbf {\bibinfo {volume} {8}},\ \bibinfo
  {pages} {011031} (\bibinfo {year} {2018})}\BibitemShut {NoStop}%
\bibitem [{\citenamefont {Ferreri}\ \emph {et~al.}(2022)\citenamefont
  {Ferreri}, \citenamefont {Pfeifer}, \citenamefont {Wilhelm}, \citenamefont
  {Hofferberth},\ and\ \citenamefont {Bruschi}}]{PhysRevA.106.033502}%
  \BibitemOpen
  \bibfield  {author} {\bibinfo {author} {\bibfnamefont {A.}~\bibnamefont
  {Ferreri}}, \bibinfo {author} {\bibfnamefont {H.}~\bibnamefont {Pfeifer}},
  \bibinfo {author} {\bibfnamefont {F.~K.}\ \bibnamefont {Wilhelm}}, \bibinfo
  {author} {\bibfnamefont {S.}~\bibnamefont {Hofferberth}},\ and\ \bibinfo
  {author} {\bibfnamefont {D.~E.}\ \bibnamefont {Bruschi}},\ }\bibfield
  {title} {\bibinfo {title} {Interplay between optomechanics and the dynamical
  casimir effect},\ }\href {https://doi.org/10.1103/PhysRevA.106.033502}
  {\bibfield  {journal} {\bibinfo  {journal} {Phys. Rev. A}\ }\textbf {\bibinfo
  {volume} {106}},\ \bibinfo {pages} {033502} (\bibinfo {year}
  {2022})}\BibitemShut {NoStop}%
\bibitem [{\citenamefont {Qin}\ \emph {et~al.}(2019)\citenamefont {Qin},
  \citenamefont {Macr\`{\i}}, \citenamefont {Miranowicz}, \citenamefont
  {Savasta},\ and\ \citenamefont {Nori}}]{PhysRevA.100.062501}%
  \BibitemOpen
  \bibfield  {author} {\bibinfo {author} {\bibfnamefont {W.}~\bibnamefont
  {Qin}}, \bibinfo {author} {\bibfnamefont {V.}~\bibnamefont {Macr\`{\i}}},
  \bibinfo {author} {\bibfnamefont {A.}~\bibnamefont {Miranowicz}}, \bibinfo
  {author} {\bibfnamefont {S.}~\bibnamefont {Savasta}},\ and\ \bibinfo {author}
  {\bibfnamefont {F.}~\bibnamefont {Nori}},\ }\bibfield  {title} {\bibinfo
  {title} {Emission of photon pairs by mechanical stimulation of the squeezed
  vacuum},\ }\href {https://doi.org/10.1103/PhysRevA.100.062501} {\bibfield
  {journal} {\bibinfo  {journal} {Phys. Rev. A}\ }\textbf {\bibinfo {volume}
  {100}},\ \bibinfo {pages} {062501} (\bibinfo {year} {2019})}\BibitemShut
  {NoStop}%
\bibitem [{\citenamefont {Dodonov}(2020)}]{dodonov_fifty_2020}%
  \BibitemOpen
  \bibfield  {author} {\bibinfo {author} {\bibfnamefont {V.}~\bibnamefont
  {Dodonov}},\ }\bibfield  {title} {\bibinfo {title} {Fifty {Years} of the
  {Dynamical} {Casimir} {Effect}},\ }\href
  {https://doi.org/10.3390/physics2010007} {\bibfield  {journal} {\bibinfo
  {journal} {Physics}\ }\textbf {\bibinfo {volume} {2}},\ \bibinfo {pages} {67}
  (\bibinfo {year} {2020})}\BibitemShut {NoStop}%
\bibitem [{\citenamefont {Bruschi}(2019)}]{bruschi_time_2019}%
  \BibitemOpen
  \bibfield  {author} {\bibinfo {author} {\bibfnamefont {D.~E.}\ \bibnamefont
  {Bruschi}},\ }\bibfield  {title} {\bibinfo {title} {Time evolution of coupled
  multimode and multiresonator optomechanical systems},\ }\href
  {https://doi.org/10.1063/1.5106409} {\bibfield  {journal} {\bibinfo
  {journal} {Journal of Mathematical Physics}\ }\textbf {\bibinfo {volume}
  {60}},\ \bibinfo {pages} {062105} (\bibinfo {year} {2019})}\BibitemShut
  {NoStop}%
\bibitem [{\citenamefont {Friis}\ \emph {et~al.}(2013)\citenamefont {Friis},
  \citenamefont {Lee},\ and\ \citenamefont {Louko}}]{Friis:Lee:2013}%
  \BibitemOpen
  \bibfield  {author} {\bibinfo {author} {\bibfnamefont {N.}~\bibnamefont
  {Friis}}, \bibinfo {author} {\bibfnamefont {A.~R.}\ \bibnamefont {Lee}},\
  and\ \bibinfo {author} {\bibfnamefont {J.}~\bibnamefont {Louko}},\ }\bibfield
   {title} {\bibinfo {title} {Scalar, spinor, and photon fields under
  relativistic cavity motion},\ }\href
  {https://doi.org/10.1103/PhysRevD.88.064028} {\bibfield  {journal} {\bibinfo
  {journal} {Phys. Rev. D}\ }\textbf {\bibinfo {volume} {88}},\ \bibinfo
  {pages} {064028} (\bibinfo {year} {2013})}\BibitemShut {NoStop}%
\bibitem [{\citenamefont {Aspelmeyer}\ \emph {et~al.}(2012)\citenamefont
  {Aspelmeyer}, \citenamefont {Meystre},\ and\ \citenamefont
  {Schwab}}]{aspelmeyer2012quantum}%
  \BibitemOpen
  \bibfield  {author} {\bibinfo {author} {\bibfnamefont {M.}~\bibnamefont
  {Aspelmeyer}}, \bibinfo {author} {\bibfnamefont {P.}~\bibnamefont
  {Meystre}},\ and\ \bibinfo {author} {\bibfnamefont {K.}~\bibnamefont
  {Schwab}},\ }\bibfield  {title} {\bibinfo {title} {Quantum optomechanics},\
  }\href
  {https://pubs.aip.org/physicstoday/article/65/7/29/414018/Quantum-optomechanicsAided-by-optical-cavitiesand}
  {\bibfield  {journal} {\bibinfo  {journal} {Physics Today}\ }\textbf
  {\bibinfo {volume} {65}},\ \bibinfo {pages} {29} (\bibinfo {year}
  {2012})}\BibitemShut {NoStop}%
\bibitem [{\citenamefont {Aspelmeyer}\ \emph {et~al.}(2014)\citenamefont
  {Aspelmeyer}, \citenamefont {Kippenberg},\ and\ \citenamefont
  {Marquardt}}]{Aspelmeyer2014}%
  \BibitemOpen
  \bibfield  {author} {\bibinfo {author} {\bibfnamefont {M.}~\bibnamefont
  {Aspelmeyer}}, \bibinfo {author} {\bibfnamefont {T.~J.}\ \bibnamefont
  {Kippenberg}},\ and\ \bibinfo {author} {\bibfnamefont {F.}~\bibnamefont
  {Marquardt}},\ }\bibfield  {title} {\bibinfo {title} {Cavity optomechanics},\
  }\href {https://doi.org/10.1103/RevModPhys.86.1391} {\bibfield  {journal}
  {\bibinfo  {journal} {Rev. Mod. Phys.}\ }\textbf {\bibinfo {volume} {86}},\
  \bibinfo {pages} {1391} (\bibinfo {year} {2014})}\BibitemShut {NoStop}%
\bibitem [{\citenamefont {O’Connell}\ \emph {et~al.}(2010)\citenamefont
  {O’Connell}, \citenamefont {Hofheinz}, \citenamefont {Ansmann},
  \citenamefont {Bialczak}, \citenamefont {Lenander}, \citenamefont {Lucero},
  \citenamefont {Neeley}, \citenamefont {Sank}, \citenamefont {Wang},
  \citenamefont {Weides} \emph {et~al.}}]{Connell2010quantum}%
  \BibitemOpen
  \bibfield  {author} {\bibinfo {author} {\bibfnamefont {A.~D.}\ \bibnamefont
  {O’Connell}}, \bibinfo {author} {\bibfnamefont {M.}~\bibnamefont
  {Hofheinz}}, \bibinfo {author} {\bibfnamefont {M.}~\bibnamefont {Ansmann}},
  \bibinfo {author} {\bibfnamefont {R.~C.}\ \bibnamefont {Bialczak}}, \bibinfo
  {author} {\bibfnamefont {M.}~\bibnamefont {Lenander}}, \bibinfo {author}
  {\bibfnamefont {E.}~\bibnamefont {Lucero}}, \bibinfo {author} {\bibfnamefont
  {M.}~\bibnamefont {Neeley}}, \bibinfo {author} {\bibfnamefont
  {D.}~\bibnamefont {Sank}}, \bibinfo {author} {\bibfnamefont {H.}~\bibnamefont
  {Wang}}, \bibinfo {author} {\bibfnamefont {M.}~\bibnamefont {Weides}}, \emph
  {et~al.},\ }\bibfield  {title} {\bibinfo {title} {Quantum ground state and
  single-phonon control of a mechanical resonator},\ }\href
  {https://www.nature.com/articles/nature08967.} {\bibfield  {journal}
  {\bibinfo  {journal} {Nature}\ }\textbf {\bibinfo {volume} {464}},\ \bibinfo
  {pages} {697} (\bibinfo {year} {2010})}\BibitemShut {NoStop}%
\bibitem [{\citenamefont {Johansson}\ \emph {et~al.}(2010)\citenamefont
  {Johansson}, \citenamefont {Johansson}, \citenamefont {Wilson},\ and\
  \citenamefont {Nori}}]{Johansson2010}%
  \BibitemOpen
  \bibfield  {author} {\bibinfo {author} {\bibfnamefont {J.~R.}\ \bibnamefont
  {Johansson}}, \bibinfo {author} {\bibfnamefont {G.}~\bibnamefont
  {Johansson}}, \bibinfo {author} {\bibfnamefont {C.~M.}\ \bibnamefont
  {Wilson}},\ and\ \bibinfo {author} {\bibfnamefont {F.}~\bibnamefont {Nori}},\
  }\bibfield  {title} {\bibinfo {title} {Dynamical {C}asimir effect in
  superconducting microwave circuits},\ }\href
  {https://doi.org/10.1103/PhysRevA.82.052509} {\bibfield  {journal} {\bibinfo
  {journal} {Phys. Rev. A}\ }\textbf {\bibinfo {volume} {82}},\ \bibinfo
  {pages} {052509} (\bibinfo {year} {2010})}\BibitemShut {NoStop}%
\bibitem [{\citenamefont {Wilson}\ \emph {et~al.}(2011)\citenamefont {Wilson},
  \citenamefont {Johansson}, \citenamefont {Pourkabirian}, \citenamefont
  {Simoen}, \citenamefont {Johansson}, \citenamefont {Duty}, \citenamefont
  {Nori},\ and\ \citenamefont {Delsing}}]{Wilson2011}%
  \BibitemOpen
  \bibfield  {author} {\bibinfo {author} {\bibfnamefont {C.}~\bibnamefont
  {Wilson}}, \bibinfo {author} {\bibfnamefont {G.}~\bibnamefont {Johansson}},
  \bibinfo {author} {\bibfnamefont {A.}~\bibnamefont {Pourkabirian}}, \bibinfo
  {author} {\bibfnamefont {M.}~\bibnamefont {Simoen}}, \bibinfo {author}
  {\bibfnamefont {J.}~\bibnamefont {Johansson}}, \bibinfo {author}
  {\bibfnamefont {T.}~\bibnamefont {Duty}}, \bibinfo {author} {\bibfnamefont
  {F.}~\bibnamefont {Nori}},\ and\ \bibinfo {author} {\bibfnamefont
  {P.}~\bibnamefont {Delsing}},\ }\bibfield  {title} {\bibinfo {title}
  {Observation of the dynamical {C}asimir effect in a superconducting
  circuit},\ }\href
  {http://www.nature.com/nature/journal/v479/n7373/abs/nature10561.html}
  {\bibfield  {journal} {\bibinfo  {journal} {Nature}\ }\textbf {\bibinfo
  {volume} {479}},\ \bibinfo {pages} {376} (\bibinfo {year}
  {2011})}\BibitemShut {NoStop}%
\bibitem [{\citenamefont {Johansson}\ \emph {et~al.}(2014)\citenamefont
  {Johansson}, \citenamefont {Johansson},\ and\ \citenamefont
  {Nori}}]{johansson_optomechanical-like_2014}%
  \BibitemOpen
  \bibfield  {author} {\bibinfo {author} {\bibfnamefont {J.~R.}\ \bibnamefont
  {Johansson}}, \bibinfo {author} {\bibfnamefont {G.}~\bibnamefont
  {Johansson}},\ and\ \bibinfo {author} {\bibfnamefont {F.}~\bibnamefont
  {Nori}},\ }\bibfield  {title} {\bibinfo {title} {Optomechanical-like coupling
  between superconducting resonators},\ }\href
  {https://doi.org/10.1103/PhysRevA.90.053833} {\bibfield  {journal} {\bibinfo
  {journal} {Physical Review A}\ }\textbf {\bibinfo {volume} {90}},\ \bibinfo
  {pages} {053833} (\bibinfo {year} {2014})}\BibitemShut {NoStop}%
\bibitem [{\citenamefont {Kim}\ \emph {et~al.}(2015)\citenamefont {Kim},
  \citenamefont {Johansson},\ and\ \citenamefont {Nori}}]{Kim2015}%
  \BibitemOpen
  \bibfield  {author} {\bibinfo {author} {\bibfnamefont {E.-j.}\ \bibnamefont
  {Kim}}, \bibinfo {author} {\bibfnamefont {J.~R.}\ \bibnamefont {Johansson}},\
  and\ \bibinfo {author} {\bibfnamefont {F.}~\bibnamefont {Nori}},\ }\bibfield
  {title} {\bibinfo {title} {Circuit analog of quadratic optomechanics},\
  }\href {https://doi.org/10.1103/PhysRevA.91.033835} {\bibfield  {journal}
  {\bibinfo  {journal} {Phys. Rev. A}\ }\textbf {\bibinfo {volume} {91}},\
  \bibinfo {pages} {033835} (\bibinfo {year} {2015})}\BibitemShut {NoStop}%
\bibitem [{\citenamefont {Wang}\ \emph {et~al.}(2023)\citenamefont {Wang},
  \citenamefont {Hu}, \citenamefont {Macr\`{\i}}, \citenamefont {Xiang},\ and\
  \citenamefont {Nori}}]{Wang2023}%
  \BibitemOpen
  \bibfield  {author} {\bibinfo {author} {\bibfnamefont {B.}~\bibnamefont
  {Wang}}, \bibinfo {author} {\bibfnamefont {J.-M.}\ \bibnamefont {Hu}},
  \bibinfo {author} {\bibfnamefont {V.}~\bibnamefont {Macr\`{\i}}}, \bibinfo
  {author} {\bibfnamefont {Z.-L.}\ \bibnamefont {Xiang}},\ and\ \bibinfo
  {author} {\bibfnamefont {F.}~\bibnamefont {Nori}},\ }\bibfield  {title}
  {\bibinfo {title} {Coherent resonant coupling between atoms and a mechanical
  oscillator mediated by cavity-vacuum fluctuations},\ }\href
  {https://doi.org/10.1103/PhysRevResearch.5.013075} {\bibfield  {journal}
  {\bibinfo  {journal} {Phys. Rev. Res.}\ }\textbf {\bibinfo {volume} {5}},\
  \bibinfo {pages} {013075} (\bibinfo {year} {2023})}\BibitemShut {NoStop}%
\bibitem [{\citenamefont {Russo}\ \emph {et~al.}(2023)\citenamefont {Russo},
  \citenamefont {Mercurio}, \citenamefont {Mauceri}, \citenamefont {Lo~Franco},
  \citenamefont {Nori}, \citenamefont {Savasta},\ and\ \citenamefont
  {Macr\`{\i}}}]{Russo2023}%
  \BibitemOpen
  \bibfield  {author} {\bibinfo {author} {\bibfnamefont {E.}~\bibnamefont
  {Russo}}, \bibinfo {author} {\bibfnamefont {A.}~\bibnamefont {Mercurio}},
  \bibinfo {author} {\bibfnamefont {F.}~\bibnamefont {Mauceri}}, \bibinfo
  {author} {\bibfnamefont {R.}~\bibnamefont {Lo~Franco}}, \bibinfo {author}
  {\bibfnamefont {F.}~\bibnamefont {Nori}}, \bibinfo {author} {\bibfnamefont
  {S.}~\bibnamefont {Savasta}},\ and\ \bibinfo {author} {\bibfnamefont
  {V.}~\bibnamefont {Macr\`{\i}}},\ }\bibfield  {title} {\bibinfo {title}
  {Optomechanical two-photon hopping},\ }\href
  {https://doi.org/10.1103/PhysRevResearch.5.013221} {\bibfield  {journal}
  {\bibinfo  {journal} {Phys. Rev. Res.}\ }\textbf {\bibinfo {volume} {5}},\
  \bibinfo {pages} {013221} (\bibinfo {year} {2023})}\BibitemShut {NoStop}%
\bibitem [{\citenamefont {Breuer}\ and\ \citenamefont
  {Petruccione}(2002)}]{breuer2002theory}%
  \BibitemOpen
  \bibfield  {author} {\bibinfo {author} {\bibfnamefont {H.-P.}\ \bibnamefont
  {Breuer}}\ and\ \bibinfo {author} {\bibfnamefont {F.}~\bibnamefont
  {Petruccione}},\ }\href@noop {} {\emph {\bibinfo {title} {The theory of open
  quantum systems}}}\ (\bibinfo  {publisher} {Oxford University Press on
  Demand},\ \bibinfo {year} {2002})\BibitemShut {NoStop}%
\bibitem [{\citenamefont {Beaudoin}\ \emph {et~al.}(2011)\citenamefont
  {Beaudoin}, \citenamefont {Gambetta},\ and\ \citenamefont
  {Blais}}]{beaudoin_dissipation_2011}%
  \BibitemOpen
  \bibfield  {author} {\bibinfo {author} {\bibfnamefont {F.}~\bibnamefont
  {Beaudoin}}, \bibinfo {author} {\bibfnamefont {J.~M.}\ \bibnamefont
  {Gambetta}},\ and\ \bibinfo {author} {\bibfnamefont {A.}~\bibnamefont
  {Blais}},\ }\bibfield  {title} {\bibinfo {title} {Dissipation and ultrastrong
  coupling in circuit {QED}},\ }\href
  {https://doi.org/10.1103/PhysRevA.84.043832} {\bibfield  {journal} {\bibinfo
  {journal} {Physical Review A}\ }\textbf {\bibinfo {volume} {84}},\ \bibinfo
  {pages} {043832} (\bibinfo {year} {2011})}\BibitemShut {NoStop}%
\bibitem [{\citenamefont {Settineri}\ \emph {et~al.}(2018)\citenamefont
  {Settineri}, \citenamefont {Macr\'{\i}}, \citenamefont {Ridolfo},
  \citenamefont {Di~Stefano}, \citenamefont {Kockum}, \citenamefont {Nori},\
  and\ \citenamefont {Savasta}}]{settineri2018}%
  \BibitemOpen
  \bibfield  {author} {\bibinfo {author} {\bibfnamefont {A.}~\bibnamefont
  {Settineri}}, \bibinfo {author} {\bibfnamefont {V.}~\bibnamefont
  {Macr\'{\i}}}, \bibinfo {author} {\bibfnamefont {A.}~\bibnamefont {Ridolfo}},
  \bibinfo {author} {\bibfnamefont {O.}~\bibnamefont {Di~Stefano}}, \bibinfo
  {author} {\bibfnamefont {A.~F.}\ \bibnamefont {Kockum}}, \bibinfo {author}
  {\bibfnamefont {F.}~\bibnamefont {Nori}},\ and\ \bibinfo {author}
  {\bibfnamefont {S.}~\bibnamefont {Savasta}},\ }\bibfield  {title} {\bibinfo
  {title} {Dissipation and thermal noise in hybrid quantum systems in the
  ultrastrong-coupling regime},\ }\href
  {https://doi.org/10.1103/PhysRevA.98.053834} {\bibfield  {journal} {\bibinfo
  {journal} {Phys. Rev. A}\ }\textbf {\bibinfo {volume} {98}},\ \bibinfo
  {pages} {053834} (\bibinfo {year} {2018})}\BibitemShut {NoStop}%
\bibitem [{\citenamefont {Sillanp{\"a}{\"a}}\ \emph {et~al.}(2007)\citenamefont
  {Sillanp{\"a}{\"a}}, \citenamefont {Park},\ and\ \citenamefont
  {Simmonds}}]{sillanpaa2007coherent}%
  \BibitemOpen
  \bibfield  {author} {\bibinfo {author} {\bibfnamefont {M.~A.}\ \bibnamefont
  {Sillanp{\"a}{\"a}}}, \bibinfo {author} {\bibfnamefont {J.~I.}\ \bibnamefont
  {Park}},\ and\ \bibinfo {author} {\bibfnamefont {R.~W.}\ \bibnamefont
  {Simmonds}},\ }\bibfield  {title} {\bibinfo {title} {Coherent quantum state
  storage and transfer between two phase qubits via a resonant cavity},\ }\href
  {https://www.nature.com/articles/nature06124} {\bibfield  {journal} {\bibinfo
   {journal} {Nature}\ }\textbf {\bibinfo {volume} {449}},\ \bibinfo {pages}
  {438} (\bibinfo {year} {2007})}\BibitemShut {NoStop}%
\bibitem [{\citenamefont {Hofheinz}\ \emph {et~al.}(2008)\citenamefont
  {Hofheinz}, \citenamefont {Weig}, \citenamefont {Ansmann}, \citenamefont
  {Bialczak}, \citenamefont {Lucero}, \citenamefont {Neeley}, \citenamefont
  {O’connell}, \citenamefont {Wang}, \citenamefont {Martinis},\ and\
  \citenamefont {Cleland}}]{hofheinz2008generation}%
  \BibitemOpen
  \bibfield  {author} {\bibinfo {author} {\bibfnamefont {M.}~\bibnamefont
  {Hofheinz}}, \bibinfo {author} {\bibfnamefont {E.}~\bibnamefont {Weig}},
  \bibinfo {author} {\bibfnamefont {M.}~\bibnamefont {Ansmann}}, \bibinfo
  {author} {\bibfnamefont {R.~C.}\ \bibnamefont {Bialczak}}, \bibinfo {author}
  {\bibfnamefont {E.}~\bibnamefont {Lucero}}, \bibinfo {author} {\bibfnamefont
  {M.}~\bibnamefont {Neeley}}, \bibinfo {author} {\bibfnamefont
  {A.}~\bibnamefont {O’connell}}, \bibinfo {author} {\bibfnamefont
  {H.}~\bibnamefont {Wang}}, \bibinfo {author} {\bibfnamefont {J.~M.}\
  \bibnamefont {Martinis}},\ and\ \bibinfo {author} {\bibfnamefont
  {A.}~\bibnamefont {Cleland}},\ }\bibfield  {title} {\bibinfo {title}
  {Generation of fock states in a superconducting quantum circuit},\ }\href
  {https://www.nature.com/articles/nature07136} {\bibfield  {journal} {\bibinfo
   {journal} {Nature}\ }\textbf {\bibinfo {volume} {454}},\ \bibinfo {pages}
  {310} (\bibinfo {year} {2008})}\BibitemShut {NoStop}%
\bibitem [{\citenamefont {Hofheinz}\ \emph {et~al.}(2009)\citenamefont
  {Hofheinz}, \citenamefont {Wang}, \citenamefont {Ansmann}, \citenamefont
  {Bialczak}, \citenamefont {Lucero}, \citenamefont {Neeley}, \citenamefont
  {O'connell}, \citenamefont {Sank}, \citenamefont {Wenner}, \citenamefont
  {Martinis} \emph {et~al.}}]{hofheinz2009synthesizing}%
  \BibitemOpen
  \bibfield  {author} {\bibinfo {author} {\bibfnamefont {M.}~\bibnamefont
  {Hofheinz}}, \bibinfo {author} {\bibfnamefont {H.}~\bibnamefont {Wang}},
  \bibinfo {author} {\bibfnamefont {M.}~\bibnamefont {Ansmann}}, \bibinfo
  {author} {\bibfnamefont {R.~C.}\ \bibnamefont {Bialczak}}, \bibinfo {author}
  {\bibfnamefont {E.}~\bibnamefont {Lucero}}, \bibinfo {author} {\bibfnamefont
  {M.}~\bibnamefont {Neeley}}, \bibinfo {author} {\bibfnamefont
  {A.}~\bibnamefont {O'connell}}, \bibinfo {author} {\bibfnamefont
  {D.}~\bibnamefont {Sank}}, \bibinfo {author} {\bibfnamefont {J.}~\bibnamefont
  {Wenner}}, \bibinfo {author} {\bibfnamefont {J.~M.}\ \bibnamefont
  {Martinis}}, \emph {et~al.},\ }\bibfield  {title} {\bibinfo {title}
  {Synthesizing arbitrary quantum states in a superconducting resonator},\
  }\href {https://www.nature.com/articles/nature08005} {\bibfield  {journal}
  {\bibinfo  {journal} {Nature}\ }\textbf {\bibinfo {volume} {459}},\ \bibinfo
  {pages} {546} (\bibinfo {year} {2009})}\BibitemShut {NoStop}%
\bibitem [{\citenamefont {Wang}\ \emph {et~al.}(2011)\citenamefont {Wang},
  \citenamefont {Mariantoni}, \citenamefont {Bialczak}, \citenamefont
  {Lenander}, \citenamefont {Lucero}, \citenamefont {Neeley}, \citenamefont
  {O'Connell}, \citenamefont {Sank}, \citenamefont {Weides}, \citenamefont
  {Wenner}, \citenamefont {Yamamoto}, \citenamefont {Yin}, \citenamefont
  {Zhao}, \citenamefont {Martinis},\ and\ \citenamefont {Cleland}}]{Wang2011}%
  \BibitemOpen
  \bibfield  {author} {\bibinfo {author} {\bibfnamefont {H.}~\bibnamefont
  {Wang}}, \bibinfo {author} {\bibfnamefont {M.}~\bibnamefont {Mariantoni}},
  \bibinfo {author} {\bibfnamefont {R.~C.}\ \bibnamefont {Bialczak}}, \bibinfo
  {author} {\bibfnamefont {M.}~\bibnamefont {Lenander}}, \bibinfo {author}
  {\bibfnamefont {E.}~\bibnamefont {Lucero}}, \bibinfo {author} {\bibfnamefont
  {M.}~\bibnamefont {Neeley}}, \bibinfo {author} {\bibfnamefont {A.~D.}\
  \bibnamefont {O'Connell}}, \bibinfo {author} {\bibfnamefont {D.}~\bibnamefont
  {Sank}}, \bibinfo {author} {\bibfnamefont {M.}~\bibnamefont {Weides}},
  \bibinfo {author} {\bibfnamefont {J.}~\bibnamefont {Wenner}}, \bibinfo
  {author} {\bibfnamefont {T.}~\bibnamefont {Yamamoto}}, \bibinfo {author}
  {\bibfnamefont {Y.}~\bibnamefont {Yin}}, \bibinfo {author} {\bibfnamefont
  {J.}~\bibnamefont {Zhao}}, \bibinfo {author} {\bibfnamefont {J.~M.}\
  \bibnamefont {Martinis}},\ and\ \bibinfo {author} {\bibfnamefont {A.~N.}\
  \bibnamefont {Cleland}},\ }\bibfield  {title} {\bibinfo {title}
  {Deterministic entanglement of photons in two superconducting microwave
  resonators},\ }\href {https://doi.org/10.1103/PhysRevLett.106.060401}
  {\bibfield  {journal} {\bibinfo  {journal} {Phys. Rev. Lett.}\ }\textbf
  {\bibinfo {volume} {106}},\ \bibinfo {pages} {060401} (\bibinfo {year}
  {2011})}\BibitemShut {NoStop}%
\bibitem [{\citenamefont {Mariantoni}\ \emph {et~al.}(2011)\citenamefont
  {Mariantoni}, \citenamefont {Wang}, \citenamefont {Bialczak}, \citenamefont
  {Lenander}, \citenamefont {Lucero}, \citenamefont {Neeley}, \citenamefont
  {O’Connell}, \citenamefont {Sank}, \citenamefont {Weides}, \citenamefont
  {Wenner} \emph {et~al.}}]{mariantoni2011photon}%
  \BibitemOpen
  \bibfield  {author} {\bibinfo {author} {\bibfnamefont {M.}~\bibnamefont
  {Mariantoni}}, \bibinfo {author} {\bibfnamefont {H.}~\bibnamefont {Wang}},
  \bibinfo {author} {\bibfnamefont {R.~C.}\ \bibnamefont {Bialczak}}, \bibinfo
  {author} {\bibfnamefont {M.}~\bibnamefont {Lenander}}, \bibinfo {author}
  {\bibfnamefont {E.}~\bibnamefont {Lucero}}, \bibinfo {author} {\bibfnamefont
  {M.}~\bibnamefont {Neeley}}, \bibinfo {author} {\bibfnamefont
  {A.}~\bibnamefont {O’Connell}}, \bibinfo {author} {\bibfnamefont
  {D.}~\bibnamefont {Sank}}, \bibinfo {author} {\bibfnamefont {M.}~\bibnamefont
  {Weides}}, \bibinfo {author} {\bibfnamefont {J.}~\bibnamefont {Wenner}},
  \emph {et~al.},\ }\bibfield  {title} {\bibinfo {title} {Photon shell game in
  three-resonator circuit quantum electrodynamics},\ }\href
  {https://www.nature.com/articles/nphys1885} {\bibfield  {journal} {\bibinfo
  {journal} {Nature Physics}\ }\textbf {\bibinfo {volume} {7}},\ \bibinfo
  {pages} {287} (\bibinfo {year} {2011})}\BibitemShut {NoStop}%
\bibitem [{\citenamefont {Kockum}\ \emph {et~al.}(2017)\citenamefont {Kockum},
  \citenamefont {Macr{\`\i}}, \citenamefont {Garziano}, \citenamefont
  {Savasta},\ and\ \citenamefont {Nori}}]{kockum2017frequency}%
  \BibitemOpen
  \bibfield  {author} {\bibinfo {author} {\bibfnamefont {A.~F.}\ \bibnamefont
  {Kockum}}, \bibinfo {author} {\bibfnamefont {V.}~\bibnamefont {Macr{\`\i}}},
  \bibinfo {author} {\bibfnamefont {L.}~\bibnamefont {Garziano}}, \bibinfo
  {author} {\bibfnamefont {S.}~\bibnamefont {Savasta}},\ and\ \bibinfo {author}
  {\bibfnamefont {F.}~\bibnamefont {Nori}},\ }\bibfield  {title} {\bibinfo
  {title} {Frequency conversion in ultrastrong cavity qed},\ }\href
  {https://www.nature.com/articles/s41598-017-04225-3} {\bibfield  {journal}
  {\bibinfo  {journal} {Scientific reports}\ }\textbf {\bibinfo {volume} {7}},\
  \bibinfo {pages} {5313} (\bibinfo {year} {2017})}\BibitemShut {NoStop}%
\bibitem [{\citenamefont {Qvarfort}\ \emph {et~al.}(2021)\citenamefont
  {Qvarfort}, \citenamefont {Vanner}, \citenamefont {Barker},\ and\
  \citenamefont {Bruschi}}]{PhysRevA.104.013501}%
  \BibitemOpen
  \bibfield  {author} {\bibinfo {author} {\bibfnamefont {S.}~\bibnamefont
  {Qvarfort}}, \bibinfo {author} {\bibfnamefont {M.~R.}\ \bibnamefont
  {Vanner}}, \bibinfo {author} {\bibfnamefont {P.~F.}\ \bibnamefont {Barker}},\
  and\ \bibinfo {author} {\bibfnamefont {D.~E.}\ \bibnamefont {Bruschi}},\
  }\bibfield  {title} {\bibinfo {title} {Master-equation treatment of nonlinear
  optomechanical systems with optical loss},\ }\href
  {https://doi.org/10.1103/PhysRevA.104.013501} {\bibfield  {journal} {\bibinfo
   {journal} {Phys. Rev. A}\ }\textbf {\bibinfo {volume} {104}},\ \bibinfo
  {pages} {013501} (\bibinfo {year} {2021})}\BibitemShut {NoStop}%
\bibitem [{\citenamefont {Schneiter}\ \emph {et~al.}(2020)\citenamefont
  {Schneiter}, \citenamefont {Qvarfort}, \citenamefont {Serafini},
  \citenamefont {Xuereb}, \citenamefont {Braun}, \citenamefont {R\"atzel},\
  and\ \citenamefont {Bruschi}}]{PhysRevA.101.033834}%
  \BibitemOpen
  \bibfield  {author} {\bibinfo {author} {\bibfnamefont {F.}~\bibnamefont
  {Schneiter}}, \bibinfo {author} {\bibfnamefont {S.}~\bibnamefont {Qvarfort}},
  \bibinfo {author} {\bibfnamefont {A.}~\bibnamefont {Serafini}}, \bibinfo
  {author} {\bibfnamefont {A.}~\bibnamefont {Xuereb}}, \bibinfo {author}
  {\bibfnamefont {D.}~\bibnamefont {Braun}}, \bibinfo {author} {\bibfnamefont
  {D.}~\bibnamefont {R\"atzel}},\ and\ \bibinfo {author} {\bibfnamefont
  {D.~E.}\ \bibnamefont {Bruschi}},\ }\bibfield  {title} {\bibinfo {title}
  {Optimal estimation with quantum optomechanical systems in the nonlinear
  regime},\ }\href {https://doi.org/10.1103/PhysRevA.101.033834} {\bibfield
  {journal} {\bibinfo  {journal} {Phys. Rev. A}\ }\textbf {\bibinfo {volume}
  {101}},\ \bibinfo {pages} {033834} (\bibinfo {year} {2020})}\BibitemShut
  {NoStop}%
\bibitem [{\citenamefont {Qvarfort}\ \emph {et~al.}(2020)\citenamefont
  {Qvarfort}, \citenamefont {Serafini}, \citenamefont {Xuereb}, \citenamefont
  {Braun}, \citenamefont {Rätzel},\ and\ \citenamefont
  {Bruschi}}]{Qvarfort_2020}%
  \BibitemOpen
  \bibfield  {author} {\bibinfo {author} {\bibfnamefont {S.}~\bibnamefont
  {Qvarfort}}, \bibinfo {author} {\bibfnamefont {A.}~\bibnamefont {Serafini}},
  \bibinfo {author} {\bibfnamefont {A.}~\bibnamefont {Xuereb}}, \bibinfo
  {author} {\bibfnamefont {D.}~\bibnamefont {Braun}}, \bibinfo {author}
  {\bibfnamefont {D.}~\bibnamefont {Rätzel}},\ and\ \bibinfo {author}
  {\bibfnamefont {D.~E.}\ \bibnamefont {Bruschi}},\ }\bibfield  {title}
  {\bibinfo {title} {Time-evolution of nonlinear optomechanical systems:
  interplay of mechanical squeezing and non-gaussianity},\ }\href
  {https://doi.org/10.1088/1751-8121/ab64d5} {\bibfield  {journal} {\bibinfo
  {journal} {Journal of Physics A: Mathematical and Theoretical}\ }\textbf
  {\bibinfo {volume} {53}},\ \bibinfo {pages} {075304} (\bibinfo {year}
  {2020})}\BibitemShut {NoStop}%
\bibitem [{\citenamefont {Quan}\ \emph {et~al.}(2005)\citenamefont {Quan},
  \citenamefont {Zhang},\ and\ \citenamefont {Sun}}]{quan_quantum_2005}%
  \BibitemOpen
  \bibfield  {author} {\bibinfo {author} {\bibfnamefont {H.~T.}\ \bibnamefont
  {Quan}}, \bibinfo {author} {\bibfnamefont {P.}~\bibnamefont {Zhang}},\ and\
  \bibinfo {author} {\bibfnamefont {C.~P.}\ \bibnamefont {Sun}},\ }\bibfield
  {title} {\bibinfo {title} {Quantum heat engine with multilevel quantum
  systems},\ }\href {https://doi.org/10.1103/PhysRevE.72.056110} {\bibfield
  {journal} {\bibinfo  {journal} {Physical Review E}\ }\textbf {\bibinfo
  {volume} {72}},\ \bibinfo {pages} {056110} (\bibinfo {year}
  {2005})}\BibitemShut {NoStop}%
\bibitem [{\citenamefont {Albash}\ \emph {et~al.}(2012)\citenamefont {Albash},
  \citenamefont {Boixo}, \citenamefont {Lidar},\ and\ \citenamefont
  {Zanardi}}]{Albash_2012}%
  \BibitemOpen
  \bibfield  {author} {\bibinfo {author} {\bibfnamefont {T.}~\bibnamefont
  {Albash}}, \bibinfo {author} {\bibfnamefont {S.}~\bibnamefont {Boixo}},
  \bibinfo {author} {\bibfnamefont {D.~A.}\ \bibnamefont {Lidar}},\ and\
  \bibinfo {author} {\bibfnamefont {P.}~\bibnamefont {Zanardi}},\ }\bibfield
  {title} {\bibinfo {title} {Quantum adiabatic markovian master equations},\
  }\href {https://doi.org/10.1088/1367-2630/14/12/123016} {\bibfield  {journal}
  {\bibinfo  {journal} {New Journal of Physics}\ }\textbf {\bibinfo {volume}
  {14}},\ \bibinfo {pages} {123016} (\bibinfo {year} {2012})}\BibitemShut
  {NoStop}%
\bibitem [{\citenamefont {Abah}\ \emph {et~al.}(2012)\citenamefont {Abah},
  \citenamefont {Roßnagel}, \citenamefont {Jacob}, \citenamefont {Deffner},
  \citenamefont {Schmidt-Kaler}, \citenamefont {Singer},\ and\ \citenamefont
  {Lutz}}]{abah_single-ion_2012}%
  \BibitemOpen
  \bibfield  {author} {\bibinfo {author} {\bibfnamefont {O.}~\bibnamefont
  {Abah}}, \bibinfo {author} {\bibfnamefont {J.}~\bibnamefont {Roßnagel}},
  \bibinfo {author} {\bibfnamefont {G.}~\bibnamefont {Jacob}}, \bibinfo
  {author} {\bibfnamefont {S.}~\bibnamefont {Deffner}}, \bibinfo {author}
  {\bibfnamefont {F.}~\bibnamefont {Schmidt-Kaler}}, \bibinfo {author}
  {\bibfnamefont {K.}~\bibnamefont {Singer}},\ and\ \bibinfo {author}
  {\bibfnamefont {E.}~\bibnamefont {Lutz}},\ }\bibfield  {title} {\bibinfo
  {title} {Single-{Ion} {Heat} {Engine} at {Maximum} {Power}},\ }\href
  {https://doi.org/10.1103/PhysRevLett.109.203006} {\bibfield  {journal}
  {\bibinfo  {journal} {Physical Review Letters}\ }\textbf {\bibinfo {volume}
  {109}},\ \bibinfo {pages} {203006} (\bibinfo {year} {2012})}\BibitemShut
  {NoStop}%
\bibitem [{\citenamefont {Roßnagel}\ \emph {et~al.}(2014)\citenamefont
  {Roßnagel}, \citenamefont {Abah}, \citenamefont {Schmidt-Kaler},
  \citenamefont {Singer},\ and\ \citenamefont
  {Lutz}}]{rosnagel_nanoscale_2014}%
  \BibitemOpen
  \bibfield  {author} {\bibinfo {author} {\bibfnamefont {J.}~\bibnamefont
  {Roßnagel}}, \bibinfo {author} {\bibfnamefont {O.}~\bibnamefont {Abah}},
  \bibinfo {author} {\bibfnamefont {F.}~\bibnamefont {Schmidt-Kaler}}, \bibinfo
  {author} {\bibfnamefont {K.}~\bibnamefont {Singer}},\ and\ \bibinfo {author}
  {\bibfnamefont {E.}~\bibnamefont {Lutz}},\ }\bibfield  {title} {\bibinfo
  {title} {Nanoscale {Heat} {Engine} {Beyond} the {Carnot} {Limit}},\ }\href
  {https://doi.org/10.1103/PhysRevLett.112.030602} {\bibfield  {journal}
  {\bibinfo  {journal} {Physical Review Letters}\ }\textbf {\bibinfo {volume}
  {112}},\ \bibinfo {pages} {030602} (\bibinfo {year} {2014})}\BibitemShut
  {NoStop}%
\bibitem [{\citenamefont {Settineri}\ \emph {et~al.}(2019)\citenamefont
  {Settineri}, \citenamefont {Macr\`{\i}}, \citenamefont {Garziano},
  \citenamefont {Di~Stefano}, \citenamefont {Nori},\ and\ \citenamefont
  {Savasta}}]{PhysRevA.100.022501}%
  \BibitemOpen
  \bibfield  {author} {\bibinfo {author} {\bibfnamefont {A.}~\bibnamefont
  {Settineri}}, \bibinfo {author} {\bibfnamefont {V.}~\bibnamefont
  {Macr\`{\i}}}, \bibinfo {author} {\bibfnamefont {L.}~\bibnamefont
  {Garziano}}, \bibinfo {author} {\bibfnamefont {O.}~\bibnamefont
  {Di~Stefano}}, \bibinfo {author} {\bibfnamefont {F.}~\bibnamefont {Nori}},\
  and\ \bibinfo {author} {\bibfnamefont {S.}~\bibnamefont {Savasta}},\
  }\bibfield  {title} {\bibinfo {title} {Conversion of mechanical noise into
  correlated photon pairs: Dynamical casimir effect from an incoherent
  mechanical drive},\ }\href {https://doi.org/10.1103/PhysRevA.100.022501}
  {\bibfield  {journal} {\bibinfo  {journal} {Phys. Rev. A}\ }\textbf {\bibinfo
  {volume} {100}},\ \bibinfo {pages} {022501} (\bibinfo {year}
  {2019})}\BibitemShut {NoStop}%
\bibitem [{\citenamefont {Crocce}\ \emph {et~al.}(2001)\citenamefont {Crocce},
  \citenamefont {Dalvit},\ and\ \citenamefont
  {Mazzitelli}}]{PhysRevA.64.013808}%
  \BibitemOpen
  \bibfield  {author} {\bibinfo {author} {\bibfnamefont {M.}~\bibnamefont
  {Crocce}}, \bibinfo {author} {\bibfnamefont {D.~A.~R.}\ \bibnamefont
  {Dalvit}},\ and\ \bibinfo {author} {\bibfnamefont {F.~D.}\ \bibnamefont
  {Mazzitelli}},\ }\bibfield  {title} {\bibinfo {title} {Resonant photon
  creation in a three-dimensional oscillating cavity},\ }\href
  {https://doi.org/10.1103/PhysRevA.64.013808} {\bibfield  {journal} {\bibinfo
  {journal} {Phys. Rev. A}\ }\textbf {\bibinfo {volume} {64}},\ \bibinfo
  {pages} {013808} (\bibinfo {year} {2001})}\BibitemShut {NoStop}%
\end{thebibliography}%
\appendix
\onecolumngrid
\newpage
\section{Derivation of the Hamiltonian}\label{appendixA:DerHam}
The literature shows different ways to derive the Hamiltonian of the system. A possible starting point could be the classical equation of motion of both the cavity and the two walls using time-dependent boundary conditions, as done by Law in \cite{law_interaction_1995}. However, in this section we show how to derive the Hamiltonian of the system from first principles employing the protocol in \cite{PhysRevA.106.033502}. This procedure is based on the idea that the position of a non-fixed wall undergoes a fluctuation described by a quantum harmonic oscillator. This concept avoids the treatment of the problem starting from dynamical equations, since a real (classical) motion of the wall would occur only in presence of coherence.

We start from the Lagrangian density of a massless scalar field in (3+1)-dimension:
\begin{align}\label{lagrangian:density}
\mathcal{L}(t,\bold x)=\frac{1}{2}\partial_\mu\phi\partial^\mu\phi.
\end{align}
The equation of motion of such Lagrangian density with static Dirichlet boundary conditions is the Klein Gordon equation $\partial_t^2\phi-\nabla^2\phi$=0, which can be solved by any scalar field of the form
\begin{align}\label{field:expression}
\phi(t,\bold x)=&\sum_{\bold n}\left[\alpha_{\bold n}\,\phi_{\bold n}(t,x)+\alpha_{\bold n}^*\,\phi^*_{\bold n}(t,x)\right],
\end{align}
with modes 
\begin{align}\label{field:modes}
\phi_{\bold n}(t,\bold x)=&\sqrt{\frac{4}{\omega_{\bold n} V}}e^{-i\omega_{\bold n}t}\sin\left(\frac{n_x\pi}{L_x}x\right)\sin\left(\frac{n_y\pi}{L_y}y\right)\sin\left(\frac{ n_z\pi}{L_z}z\right),
\end{align}
where the dispersion law reads 
\begin{align}\label{field:disp}
\omega_{\bold n}:=\sqrt{\left(\frac{n_x\pi}{L_x}\right)^2+\left(\frac{n_y\pi}{L_y}\right)^2+\left(\frac{n_z\pi}{L_z}\right)^2},
\end{align}
with $\bold n\equiv(n_x,n_y,n_y)$ a set of positive integer numbers, and box having volume $V=L_x L_y L_z$ \cite{PhysRevA.64.013808}. With the typical formalism, well-known from field theory, we can calculate the Hamiltonian density $\mathcal{H}(t,\bold x)=\frac{1}{2}\left[\Pi^2(t,\bold x)+(\nabla\phi(t,\bold x))^2\right]$, with canonical momentum $\Pi(t,\bold x):=-\partial_t\phi(t,\bold x)$.

Hence, the protocol consists in four steps: \begin{itemize}
    \item[i)] Extension of the box length with respect to $L_x$: $L_x\rightarrow L_x+\Delta L_x$, where $\Delta L_x/L_x\ll1$, and Taylor-expansion of $\mathcal{H}(t,\bold x)$ up to the first order in $\Delta L_x/L_x$;
    \item[ii)] Spatial integration of the Hamiltonian density in volume $V$, thereby obtaining the classical Hamiltonian $H:=\int \mathcal{H}\, dV$;
    \item[iii)] Quantization of the field Fourier coefficients $\alpha_{\bold n}$:
\begin{align}\label{hamiltonian:explicit}
\alpha_{\bold n}\rightarrow&\hat{a}_{\bold n}\nonumber,\\
\alpha_{\bold n}^*\rightarrow&\hat{a}^\dag_{\bold n}\nonumber,
\end{align}
which now fulfill the standard commutation rules $[\hat a_{\bold n},\hat a_{{\bold n}'}^\dag]=\delta_{{\bold n}{\bold n}'}$.
We now need to quantize also the position of the two walls. If we assume that our system has a cylindrical symmetry along the $x$-axis and that in our frame of reference the wall 1 is always at $x=0$, the second wall undergoes a fluctuation characterized by two Fourier harmonics: $\Delta L_x=\Delta L_1+\Delta L_2$. Such harmonic fluctuations do not interact with each other and can be treated as two independent harmonic oscillators. Therefore, the quantization of such degrees of freedom leads to:
\begin{align}
\Delta L_1\rightarrow& \delta L_1 (\hat{b}_1^\dag+\hat{b}_1),\\
\Delta L_2\rightarrow& \delta L_2 (\hat{b}_2^\dag+\hat{b}_2),
\end{align}
where we introduced the annihilation and creation operators of two quantum harmonic oscillators fulfilling the standard commutation relations: $[\hat{b}_1,\hat{b}_1^\dag]=[\hat{b}_2,\hat{b}_2^\dag]=1$, while all other commutators vanish. We notice that $\delta L_1$ and $\delta L_2$ are the zero-point fluctuations of two harmonic oscillators having different masses \cite{Aspelmeyer2014}.
\item[iv)] As last step, we rewrite the quantum Hamiltonian $\hat{H}$ in normal order and we introduce two dimensionless amplitudes $\epsilon_1:=\delta L_1/L_x\ll1$ and $\epsilon_2:=\delta L_2/L_x\ll1$.
\end{itemize}

In the end, this procedure yields the Hamiltonian
\begin{align}\label{quantum:hamiltonian:explicit}
\hat{H}(t)=&\hat H_0+\hat H_{\textrm{I}},
\end{align}
where each term reads
\begin{align}\label{quantum:hamiltonian:terms2}
\hat{H}_0:=&\sum_{\bold n}\omega_{\bold n}\,\hat{a}_{\bold n}^\dag\hat{a}_{\bold n}+\omega_1\hat{b}_1^\dag\hat{b}_1+\omega_2\hat{b}_2^\dag\hat{b}_2\nonumber,\\
\hat{H}_{\textrm{I}}:=&2\sum_{\bold n}\left[\frac{k(n_y)^2+k(n_z)^2}{\omega_{\bold n}}\hat a_{\bold n}^\dag\hat a_{\bold n} -4\sum_{m_x}(-1)^{n_x+m_x}\frac{k(n_x)k(m_x)}{\sqrt{\omega_{\bold n}\,\omega_{m_x,n_y,n_z}}} \hat X_{\bold n}\hat X_{m_x,n_y,n_z}\right](\epsilon_1\hat X_{\textrm{b}1}+\epsilon_2\hat X_{\textrm{b}2}).
\end{align}
In order to simplify the notation, we introduced  the wave vector $ k(n_w)=\pi n_w/L_w$, with $w=x,y,z$, the quadrature position operators $\hat X_{\textrm{b},j}=\frac{1}{2}(\hat b_j^\dagger+\hat b_j)$, with $j=1,2$, and $\hat X_{\bold n}=\frac{1}{2}(\hat a_{\bold n}^\dagger+\hat a_{\bold n})$. We note that the ambiguity on the negative sign in $\hat{H}_{\textrm{I}}$ is solved by including all terms of the Taylor expansion with respect to $\delta L$ \cite{law_interaction_1995}.

For our purposes, such Hamiltonian can be drastically simplified by assuming:
\begin{enumerate}
    \item {The constraint $L_x\gg L_y,L_z$ on the magnitude of the length of the edges of the piston. This is motivated by the fact that there are no excitations initially present in the $y$ and $z$ degrees of freedom. Given the higher energy required to excite these degrees of freedom (since the energy gaps are inversely proportional to the corresponding length) it is reasonable to assume that they will remain unexcited. Thus, we can drop the first term in the square brackets and obtain Hamiltonian (1) in the main text. }
    \item {The presence of only one cavity mode, say the fundamental mode $\omega_{1,1,1}=\omega_{\textrm{c}}$. This is a reasonable assumption, since other cavity modes with $n_x>1$ are initialized in the vacuum state.}
\end{enumerate} 
Under such assumptions, the system Hamiltonian is reduced to Eq.~(1) of the main text.

\section{The master equation in the dressed picture and the dressed operators}\label{te:appendix}

In this section we want to review the necessary tools to describe the time evolution of the quantities of interest. Working in the Schrödinger picture, we first need to solve the master equation for the density operator; however, since we are dealing with a strongly interacting system, the best way to proceed is to employ the master equation in the dressed picture \cite{beaudoin_dissipation_2011}. To do this, first
let us introduce the transition amplitudes for the canonical position operators calculated on the dressed basis: $c_{ij}=\langle i\lvert \hat a+\hat a^\dag\rvert j\rangle$, $u_{ij}=\langle i\lvert \hat b_1+\hat b_1^\dag\rvert j\rangle$ and $v_{ij}=\langle i\lvert \hat b_2+\hat b_2^\dag\rvert j\rangle$, where the state $\rvert i\rangle$ is the $i$-th eigenstate of the Hamiltonian with eigenenergy $E_i$. 

We assume that the three subsystems are coupled to three different baths: in particular, the wall 1 is coupled to a cold bath with damping rate $\gamma_1$, the wall 2 is coupled to a hot bath with damping rate $\gamma_2$. Finally, we also assume that the cavity always weakly interacts with its own bath at $T\simeq 0$, with damping rate $\kappa\simeq 0$. Although it does not play an active role in our dynamics, the interaction with the third bath whose temperature is not exactly zero is expected in experimental scenarios and prevents the violation of the third law of thermodynamics.

The master equation in the dressed picture has been obtained before \cite{beaudoin_dissipation_2011,settineri2018}, and it reads
\begin{equation}
\frac{d\hat{\rho}}{dt}=-i[\hat H,\hat{\rho}]+(\hat{\mathcal{L}}_c+\hat{\mathcal{L}}_{u}+\hat{\mathcal{L}}_{v})\hat{\rho},
\label{me:appendix}
\end{equation}
where we have defined
\begin{align}
    \hat{\mathcal{L}}_x\hat{\rho}=y\sum_{j,i>j}\lvert x_{ij}\rvert^2\left\{n_{ij}(T)\mathcal{D}[\hat P_{ij}]\hat{\rho}+(1+n_{ij}(T))\mathcal{D}[\hat P_{ji}]\hat{\rho}\right\}
\end{align}
that needs to be supplemented by the quantities $x=c, u, v$, the rates of losses $y=\gamma_1,\gamma_2,\kappa$, the thermal excitation numbers $n_{ij}(T)=(e^{(E_i-E_j)/T}-1)^{-1}$, the superoperators 
\begin{align}
\mathcal{D}[\hat P_{ij}]\hat{\rho}=\frac{1}{2}\bigl(2\hat P_{ij}\hat{\rho}\hat P_{ij}^\dag-\hat{\rho}\hat P_{ij}^\dag\hat P_{ij}-\hat P_{ij}^\dag\hat P_{ij}\hat{\rho}\bigr),
\end{align}
and the transition operators $\hat P_{ij}=\lvert i\rangle\langle j\rvert$.

In order to evaluate the quantity of interest, it is necessary to define a set of dressed annihilation operators for the various subsystems, and we have 
\begin{align}
\hat A=&\sum_{j,i>j}c_{ij}\hat P_{ij},\nonumber\\
\hat B_1=&\sum_{j,i>j}u_{ij}\hat P_{ij},\nonumber\\
\hat B_2=&\sum_{j,i>j}v_{ij}\hat P_{ij}.
\label{dressop}
\end{align}
We recall that the photon as well as the phonon population at any instant $t$ of time is given by $N_{c}(t)=\textrm{Tr}[\hat A^\dag\hat A\rho(t)]$ and $N_{\textrm{w}_{j}}(t)=\textrm{Tr}[\hat B_{j}^\dag\hat B_{j}\rho(t)]$ ($j=1,2$), respectively.

\section{The external drive}
The total Hamiltonian includes a time-dependent term acting as an external drive that periodically shifts the effective frequency of cavity mode from $\tilde\omega_{\textrm{c},1}$ to $\tilde\omega_{\textrm{c},2}$ and backwards. From a physical perspective, the role of such external drive is to simulate the compression and the expansion of the cavity, thereby activating the resonant phonon-photon conversion with either $\tilde\omega_{\textrm{c},1}$ or $\tilde\omega_{\textrm{c},2}$. The explicit form of the time-dependent term is: $\hat H_{\textrm{dr}}(t)=f(t) \Delta\omega\hat A^\dag\hat A$, where $\Delta\omega=\tilde\omega_{\textrm{c},2}-\tilde\omega_{\textrm{c},1}$, and $\hat A,\hat A^\dag$ are the cavity dressed operators in \eqref{dressop}. 
The function $f(t)$ is a periodic smooth step function which allows us to rapidly vary the frequency of the cavity during the adiabatic transformations. Its explicit form is
  \begin{align}
f(t)=&\sum_i^N\left.\{\sin^2[\Omega(t-t_i)]\theta[t-t_i]\right.+\sin^2\left[\Omega\left(t-t_i-\tau\right)\right]\theta\left[t-t_i-\tau\right]\nonumber\\
&-\sin^2\left[\Omega\left(t-t_i-\Delta t\right)\right]\theta\left[t-t_i-\Delta t\right]\left.-\sin^2\left[\Omega\left(t-t_i-\tau-\Delta t\right)\right]\theta\left[t-t_i-\tau-\Delta t\right]\right\},
\end{align}
where $t_i$ indicate the instants of time when the cycle is run, $\Delta t$ is the duration of the hot isochoric, $\tau$ is the duration of the two adiabatics, and $\Omega=2\pi/\tau$. $\theta[t]$ is the Heaviside function. The duration of the cold isochoric was estimated such that $\Delta \mathcal{U}$ reaches zero at the end of the transformation (see Fig.~3 of the main text), and it corresponds to $t_2-t_1-2\tau-\Delta t$, whereas the duration of one cycle is $t_2-t_1$.

\end{document}